\documentclass{article}

\usepackage{arxiv}

\usepackage[utf8]{inputenc} 
\usepackage[T1]{fontenc}    
\usepackage{hyperref}       
\usepackage{url}            
\usepackage{booktabs}       
\usepackage{amsfonts}       
\usepackage{nicefrac}       
\usepackage{microtype}      
\usepackage{lipsum}		
\usepackage{graphicx}
\usepackage{natbib}
\setcitestyle{authoryear,open={(},close={)}}
\usepackage{doi}

\usepackage{lineno}
\usepackage[export]{adjustbox}  
\usepackage{tabularx}
\usepackage{array}  
\usepackage{bm}
\usepackage[onehalfspacing]{setspace}
\usepackage{amsmath}
\title{Beyond Flickering: Introducing Code-Modulated Motion Visual Evoked Potentials for Brain-Computer Interfacing}

\author{Hanneke A. Scheppink\thanks{Department of Informatics and Data Science, Doctoral School for Applied Research in North Rhine-Westphalia, Bochum, Germany} \\
	Faculty of Technology and Bionics\\
	Rhine-Waal University of Applied Sciences\\
	Kleve, Germany \\
	\And
    Rainer Herpers$^{*}$ \\
	Institute of Visual Computing and Graduate Institute\\
    Bonn-Rhein-Sieg University of Applied Science\\
	Sankt Augustin, Germany \\
	\And
	Jordy Thielen \\
	Radboud University \\
    Donders Institute for Brain, Cognition and Behaviour \\ 
	Nijmegen, The Netherlands \\
	\And
	Ivan Volosyak$^{*}$ \\
	Faculty of Technology and Bionics\\
	Rhine-Waal University of Applied Sciences\\
	Kleve, Germany \\
	\texttt{ivan.volosyak@hochschule-rhein-waal.de} \\
}

\date{}

\renewcommand{\shorttitle}{Code-Modulated Motion VEPs for BCI}

\hypersetup{
pdftitle={Code-Modulated Motion VEPs for BCI},
pdfsubject={q-bio.NC},
pdfauthor={Hanneke A. Scheppink, Rainer Herpers, Jordy Thielen, Ivan Volosyak},
pdfkeywords={brain-computer interface (BCI), code-modulated motion visual evoked potential (c-MVEP), code-modulated visual evoked potential (c-VEP), electroencephalography (EEG), BCI user comfort, flicker-free VEP, steady-state visual evoked potential (SSVEP), steady-state motion visual evoked potential (SSMVEP)},
colorlinks = true,linkcolor  = black,citecolor  = black,urlcolor=black
}

\begin{document}
\maketitle

\begin{abstract}
This study presents a novel code-modulated motion visual evoked potential (c-MVEP) paradigm for brain-computer interfacing (BCI). To avoid the visual discomfort and fatigue often associated with traditional flickering stimuli, this paradigm uses pseudo-random sequences to visually stimulate objects using motion. We conducted offline and online experiments, to investigate signal characteristics and evaluate practical feasibility, respectively.
In the offline experiment, EEG data were recorded and compared during sequential stimulation of a single target under four conditions: c-MVEP, code-modulated visual evoked potential (c-VEP), steady-state motion visual evoked potential (SSMVEP), and steady-state visual evoked potential (SSVEP). 
The c-MVEP evoked similar temporal and broadband spectral responses as c-VEP, with a comparable signal-to-noise ratio (SNR), although c-MVEP responses were more focused in the lower frequency range. While SSMVEP and SSVEP both showed clear harmonic oscillations, SSVEP yielded higher SNRs. Spatially, both motion-based stimulations peaked at Oz but spread across multiple electrodes, whereas both flicker-based stimulations were more localized at Oz.
In the online experiment, we evaluated a four-target BCI using the same four conditions, testing the practical feasibility of the c-MVEP paradigm. 
The c-MVEP BCI reached a mean accuracy of $85.67\,\%$ with an average selection time of $2.61$\,s, which was significantly lower than c-VEP ($97.81\,\%$; $1.15$\,s) and SSVEP ($93.42\,\%$; $1.94$\,s), but significantly higher than SSMVEP ($64.91\,\%$; $4.18$\,s). 
The subjective ratings revealed no clear preference between the motion- and flicker-based paradigms, indicating comparable user comfort.
Overall, this study demonstrates the strong potential of the newly proposed c-MVEP paradigm. By providing an effective, non-flickering alternative to traditional c-VEP and SSVEP, c-MVEP offers a highly viable approach for user-friendly BCI applications. 
\end{abstract}


\section{Introduction}
A brain-computer interface (BCI) is a system that allows real-time communication between the human brain and external devices, bypassing muscular involvement~\citep{wolpaw2012}. This makes BCI systems a promising tool for users with impaired or absent muscular function. The neural activity is most commonly measured non-invasively using electroencephalography (EEG)~\citep{wolpaw2002}. An example application of BCI is to enable communication or control through a BCI application, such as a speller, using external stimuli~\citep{rezeika2018}.

One of the visual stimulation paradigms that achieves a high information transfer rate (ITR) is the code-modulated visual evoked potential (c-VEP)~\citep{martinez-cagigal2021, miao2024}. In a c-VEP-based BCI paradigm, stimuli flicker according to a pseudo-random sequence. Often, a binary sequence is used, producing high-contrast flickering between black and white. A similar and high-performing paradigm is the steady-state visual evoked potential (SSVEP) in which stimuli flicker according to a specific and constant frequency~\citep{vialatte2010}. 

Although c-VEP and SSVEP are high-performing paradigms, a major limitation of both is the visual and mental fatigue, as well as eye strain, that can result from the high-contrast flickering of the visual stimuli~\citep{moghadam2023}. 

To address this issue, the steady-state motion visual evoked potential (SSMVEP) paradigm was proposed as an alternative for BCI applications~\citep{xie2012, xie2016}. In SSMVEP, stimuli are modulated through continuous movement rather than flickering, according to a specific frequency. This movement can be implemented, for instance, by radially zooming with periodic expansion and contraction of the stimulus size~\citep{chai2019, chai2020}, or through dynamic patterns such as expanding and contracting ring-shaped checkerboards~\citep{han2017, yan2017} or Newton rings~\citep{xie2012, xie2016}. The motion typically follows a sinusoidal modulation pattern, where the direction of movement reverses twice per cycle, a rate commonly referred to as the flip-frequency or motion-reversal frequency. The neural responses evoked by attending to these motion-based stimuli resemble those of SSVEPs, although with a weaker activation level. SSMVEP evokes responses across more channels than SSVEP, with greater activation in middle temporal areas, whereas SSVEP was more focused in occipital areas~\citep{han2018a}. 

Although SSMVEP relies on motion stimulation, it should not be mistaken for the motion-onset VEP (m-VEP), which is based on the anticipation of a single moving stimulus, typically presented in isolation at a constant speed. In contrast, in SSMVEP, all stimuli move simultaneously, each at a distinct frequency.
The design of SSMVEP allows for shorter stimulation durations and greater flexibility in the number of targets compared to m-VEP, where ITR is limited due to longer stimulation durations and a restricted number of targets~\citep{guo2008, li2017, ma2017a}. Moreover, SSMVEP avoids the direction-specific adaptation effects, commonly arising in m-VEP, that can potentially cause visual fatigue and a reduced signal-to-noise ratio (SNR)~\citep{xie2012}.

Importantly, SSMVEP is particularly well performing under conditions of high visual and mental fatigue, especially compared to SSVEP~\citep{xie2016}. However, despite this advantage, its overall performance substantially lags behind that of SSVEP and c-VEP~\citep{volosyak2020}. Additionally, SSMVEP exhibits large inter-subject variability, with some users unable to achieve effective control, whereas c-VEP has shown to provide reliable and high performance for all users~\citep{volosyak2020, martinez-cagigal2021, thielen2025}.

This study introduces the code-modulated motion visual evoked potential (c-MVEP), a novel paradigm that integrates the core principle of the established and reliable c-VEP protocol with the comfortable, low-fatigue characteristics of SSMVEP.
In this paradigm, we propose modulating the stimuli using motion following a pseudo-random sequence, rather than flickers, to evoke a code-modulated response. We hypothesize that this novel motion-based stimulation will improve user comfort~\citep{xie2012, xie2016} compared to the established c-VEP. Furthermore, by combining motion-based stimulation with code-modulation, we hypothesize that c-MVEP will allow for more reliable control than the established frequency-based motion of SSMVEP.
The current study therefore aims to evaluate the subjective user comfort of c-MVEP using a questionnaire and to assess its feasibility and applicability for BCI applications, through an online BCI implementation.

To fully investigate the proposed c-MVEP paradigm, this study uses a two-by-two factorial design to compare four paradigms, the novel c-MVEP, and the established c-VEP, SSMVEP, and SSVEP. The rationale for this design is twofold. First, we compare motion to flickering, specifically, c-MVEP against c-VEP, and SSMVEP against SSVEP, to assess how replacing flickering with motion affects comfort, signal characteristics, and performance. Second, we compare code-modulated to steady-state stimulation, specifically, c-MVEP against SSMVEP, and c-VEP against SSVEP, to verify if the superior decoding performance typically observed for c-VEP, also applies to motion-based BCIs.
This comprehensive comparison is performed using two experiments, an offline experiment to characterize and compare the neural responses of each paradigm, and an online experiment to evaluate the practical decoding performance in a BCI context. For both experiments, the subjective ratings are also collected.

\section{Materials and Methods}

\subsection{Participants}
A total of 28 subjects participated in this study. All subjects had normal or corrected-to-normal vision and reported no history of photosensitive epilepsy. One participant took part in both the offline and online experiments. This overlap was not selected based on experimental performance, as the offline experiment was not designed as a BCI experiment and both experiments were completed before any results were available. The offline experiment was conducted as an exploratory study and therefore included a relatively small sample size of 9 participants (2M, 7F, average age of $28.6 \pm 4.0$ years, between $24$-$35$ years).

In the online experiment, 20 subjects participated. However, one participant was excluded due to observable task disengagement during the online experiment.
A total of 19 participants (14M, 5F, average age of $23.7 \pm 3.9$ years, between $19$-$35$ years) were analyzed for the online experiment, of which three participants reported having previous experience with a BCI system, consistent with recruitment via voluntary participation.

Both the offline and online experiments followed a within-subject design. Each participant completed all conditions within the respective experiment.

Prior to the experiment, all participants received detailed information about the experimental procedure and any possible risks involved, all had the opportunity to opt out at any time during the experiment. Those who agreed to participate in the study signed a consent form and were financially compensated after their participation. The study and data handling were approved by the ethical committee of the medical faculty of Duisburg-Essen University (24-11957-BO), and were conducted in accordance with the Helsinki Declaration. 

\subsection{Hardware and Software}
Both the offline and online experiments were done in the same environment under comparable conditions using the same hardware. 

The EEG data were acquired with a g.USBamp amplifier (g.tec, Schiedlberg, Austria) using sixteen gel-based passive scalp electrodes placed according to the international 10-5 system at the following positions: $P_7$, $P_3$, $P_z$, $P_4$, $P_8$, $PO_7$, $PO_3$, $PO_z$, $PO_4$, $PO_8$, $O_1$, $O_z$, $O_2$, $O_9$, $I_z$, and $O_{10}$. These electrode positions were selected to optimize the acquisition of visual evoked potentials (VEPs) over the parieto-occipital cortex, see~\citet{ahmadi2019} for a larger analysis of electrode positioning for c-VEP studies.  The reference and ground electrodes were placed at $C_z$ and $AF_z$, respectively. The EEG signals were recorded with a sampling frequency $F_s$ of $600$\,Hz. Applying an abrasive electrode gel reduced the impedance to below $5$\,k$\Omega$.

The EEG data from the offline experiment were digitally filtered and epoched in software. A notch filter was applied to remove $50$\,Hz power line noise. Then, the raw EEG data were bandpass filtered from $2.0$ to $40.0$\,Hz using a fourth-order, zero-phase, infinite-impulse-response (IIR) Butterworth filter, baseline-corrected, and sliced into trials of the trial length. 
During the online experiment, the incoming continuous EEG data stream was causally filtered in real time, using a software-implemented, stateful fourth-order Butterworth bandpass filter ($2.0$ to $40.0$\,Hz). To ensure consistency, the data used to train the classifier were processed with this same digital filtering procedure.

Both experiments were conducted on a Dell Precision desktop computer equipped with an Intel Core i9 processor ($3.70$\,GHz) and an NVIDIA RTX 3070 graphics card, running Microsoft Windows 11 Education. Visual stimuli were presented on an Acer Nitro XV252Q F liquid-crystal display (1920 × 1080 pixels) with a refresh rate of $360$\,Hz. 

The stimulation of the experiments was implemented using the Unity game engine (Unity Technologies, San Francisco, CA, USA; version 6000.0.31f1). Experiment stimulation markers were sent via the Lab Streaming Layer (LSL) using LSL4Unity (v1.16.0).
Python LSL markers, such as classifier-related markers, were streamed using pyLSL (v1.17.6). EEG signals and all event markers were recorded using LabRecorder (v1.16.5). 
Data processing and analysis were conducted in Python (v3.13.7), using MNE (v1.10.2), NumPy (v2.3.3), scikit-learn (v1.7.1), and SciPy (v1.16.1). Recorded data were imported using pyxdf (v1.17.1), and classification procedures were performed using PyNTBCI (v1.8.3).

\subsection{Stimulus Generation}\label{stimGen}
This study included four types of stimulation. Specifically, the newly proposed c-MVEP with code-modulated motion stimulation, and its flickering counterpart, the code-modulated flickering c-VEP. Both used the same pseudo-random sequence, a smoothed version (c-MVEP) and the original binary sequence (c-VEP). Additionally, we implemented the frequency-based motion stimulation of SSMVEP and its frequency-based flickering counterpart, SSVEP. 

The motion conditions, c-MVEP and SSMVEP, used a radial zooming of the stimulus. This stimulation motion was previously shown to be preferred over expansion-contraction Newton rings for SSMVEP~\citep{chai2019}, and was implemented effectively in~\citet{chai2019,chai2020}. Radial zooming has also been effectively applied in hybrid stimulation, together with flickering, by~\citet{li2023d, kwon2022}. The flickering of c-VEP and SSVEP was implemented as a binary flickering between black and white.

\begin{figure}[h]
\centering

\includegraphics[width=\textwidth]{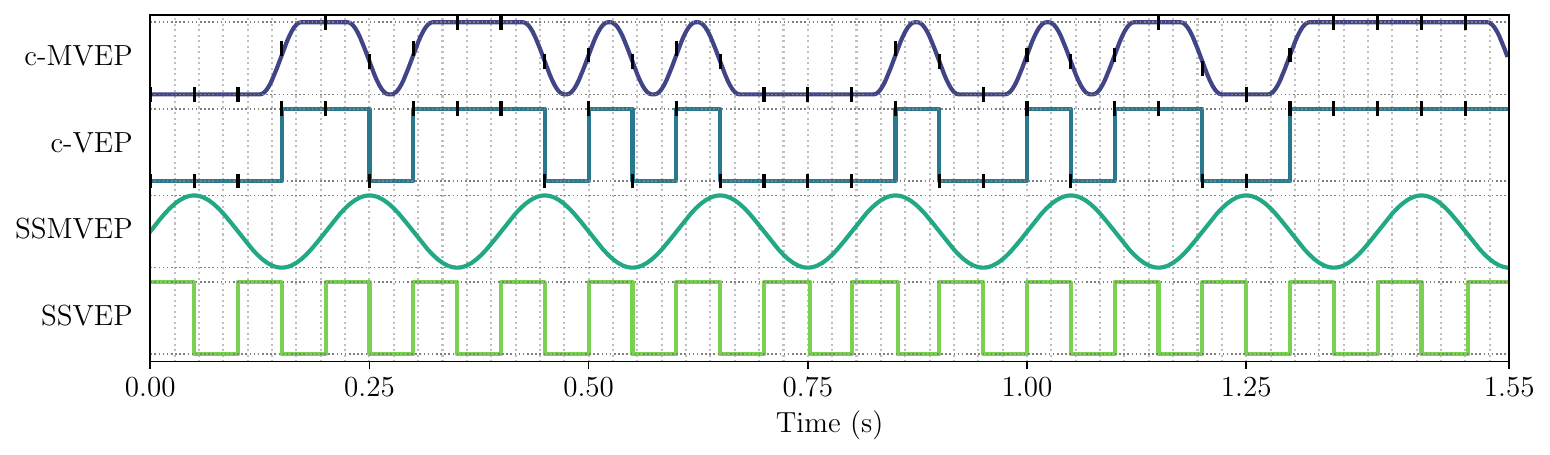}
\caption{\textbf{Stimulation waveforms from the offline experiment.} The different waveforms used in the offline experiment for the four conditions: c-MVEP, c-VEP, SSMVEP, and SSVEP. The dotted vertical lines represent each 10th frame, at a monitor refresh rate of $360$\,Hz. The black ticks in the c-MVEP and c-VEP conditions represent individual bits, of a presentation rate of $20$\,Hz, meaning 18 frames per bit. SSMVEP has a flip-frequency of $10$\,Hz, meaning a sine wave of $5$\,Hz was used, SSVEP was presented at $10$\,Hz. }\label{fig:waveforms_conditions}
\end{figure}

\begin{figure}[h]
\centering

\includegraphics[width=\textwidth]{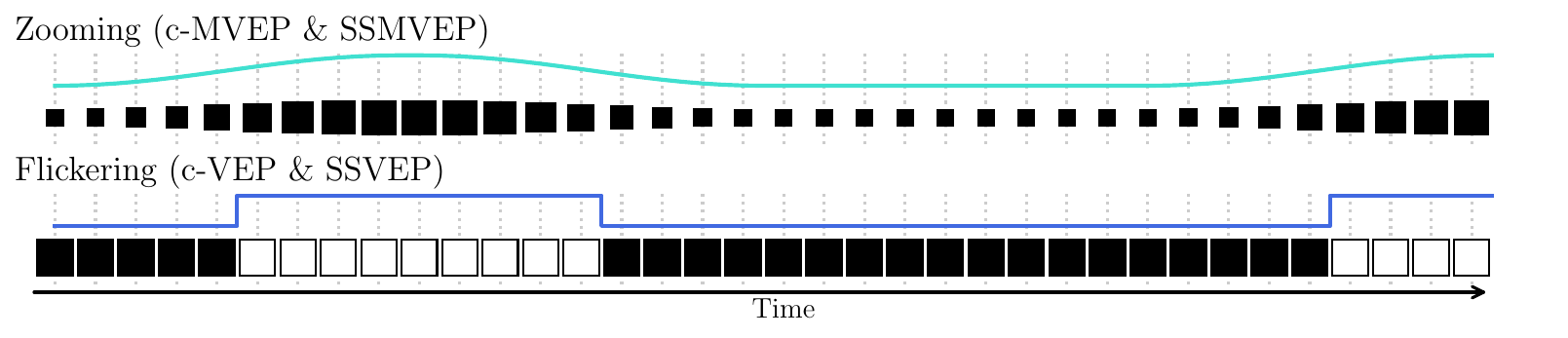}
\caption{\textbf{Zooming and flickering stimulation.} Partial waveforms of the zooming conditions (c-MVEP and SSMVEP) and the corresponding stimulus size modulation. Similarly, for the flickering conditions (c-VEP and SSVEP), a partial waveform is shown, together with the stimuli that are changing between black and white depending on the state of the waveform.}\label{fig:stim_and_waveform}
\end{figure}

To implement the zooming in c-MVEP, we used a modified pseudo-random sequence, and the original binary pseudo-random sequence for flickering in c-VEP. SSMVEP and SSVEP used distinct frequencies for stimulation. \autoref{fig:waveforms_conditions} shows a visual representation of the waveforms underlying the stimulation for each of the conditions. \autoref{fig:stim_and_waveform} shows how these waveforms were translated into motion and flickering.


\subsubsection{Code-Modulated Stimulation for c-MVEP and c-VEP}
To visually stimulate the targets in a code-modulated-based BCI system, the stimuli flicker or move according to pseudo-random sequences. For c-VEP, this is often done using a binary maximum length sequence (m-sequence)~\citep{thielen2025}, which is generated by a linear feedback shift register (LFSR). 
An m-sequence has great autocorrelation properties, specifically, a correlation of $1$ for no shift and $-1/K$ for all other shifts with $K$ the length of the m-sequence~\citep{golomb1982, meel1999, holmes2007}. In c-VEP, this binary code represents the black (`0' in the code) and white (`1' in the code) flickering.

Here, we created the initial m-sequence $c_0$ with feedback taps at register positions 5 and 2, corresponding to the primitive polynomial of $p(x) = x^5 + x^2 + 1$. The initial register state was set to $\mathbf{r}(0) = (1,1,1,1,1)$. This produced the following 31-bit m-sequence $c_0$ for the offline experiment:
\begin{equation}
    c_0 = 0001101110101000010010110011111.
\end{equation}

For the online experiment, we set four m-sequences such that the sequence started with a 1, creating a smooth start of the trial for the c-MVEP condition. 
Specifically, we right-circularly shifted the initial sequence $c_0$ by 1 bit, to create $c_1$. The second sequence $c_2$ was a $3$ bit right-circularly shifted version of $c_0$. The third sequence $c_3$ shifted $c_0$ by $5$ bits, and $c_4$ by $8$ bits.

In their original form, these binary sequences were used to directly modulate the luminance of the stimuli for c-VEP, mapping the binary states to black (`0') and white (`1'). However, directly applying a binary (square-wave) modulation to motion parameters for c-MVEP, specifically the zooming of a stimulus between a predefined minimum (`0') and maximum (`1') size, would result in abrupt transitions. Such discontinuities could lead to a perceptually rough motion and may even be experienced as flickering rather than a smooth and fluid motion. To address this, the binary m-sequences were transformed into continuous-valued sequences for c-MVEP stimulation. By smoothing the abrupt rising and falling edges of the m-sequence, similar to the approach from a code-modulated auditory evoked potential (c-AEP) study~\citep{scheppink2024}, smooth transitions between the extremes were achieved. This ensured a fluid motion while retaining the event durations and overall timing of the original m-sequence.

Specifically, we smoothed the edges using a raised sine function. First, the original m-sequence was upsampled by a factor of $F_b$, corresponding to the number of display frames per bit, here $F_b = 18$, derived from the monitor's refresh rate of $360$\,Hz and the $20$\,Hz presentation-rate of the m-sequence.
Each transition between consecutive bits was identified. For each rising edge (0 $\rightarrow$ 1) and falling edge (1 $\rightarrow$ 0), a transition window of length $L = \alpha F_b$ was placed around each detected transition, where $\alpha$ denotes a smoothing factor. Here, $\alpha = 1.0$, resulting in a transition window equal to one bit duration.
Within each transition window, indexed by $t = 0, \dots, L$, the sharp edge was replaced by a smooth sinusoidal function,
\begin{equation}
    s(t) = \dfrac{1}{2} \left( 1 + \sin\left( \pi \left( \dfrac{t}{L} - \dfrac{1}{2}\right) \right) \right) , \quad t = 0, \dots, L
\end{equation}
which smoothly connected 0 and 1. Rising edges were represented by $s(t)$, whereas the falling edges used the reversed function $1-s(t)$.

The m-sequence was only modified within the transition windows, outside of these windows, the sequence remained unchanged. This created a smooth motion modulator, without introducing temporal artifacts or distorting the properties of the m-sequence, see \autoref{fig:waveforms_conditions}. The correlation properties of the m-sequence were minimally changed, as shown in Figure S1 of the Supplementary Material.

Concisely, to translate a binary m-sequence into a fluid and continuous motion, the abrupt transitions were first smoothed. This smoothed m-sequence, which ranged from 0 to 1, was then linearly mapped to a restricted stimulus size. Specifically, the stimulus size was limited to a minimum of $50\,\%$ (for a 0 in the code) and a maximum of $100\,\%$ of its original size (for a 1 in the code). This ensured that the stimulus zoomed between half and its full original size, see~\autoref{fig:stim_and_waveform} and \autoref{fig:exp_protocol}. We opted for a minimum restriction to $50\,\%$ due to the limited number of frames available, here 18 frames, to transition between the extremes. All intermediate values in the smoothed sequence were mapped to a corresponding zoom scale, for instance, a value of 0.75 in the sequence translated to a stimulus size of $87.5\,\%$, a value of 0.5 to $75\,\%$, and a value of 0.25 to $62.5\,\%$, as can be seen in ~\autoref{fig:stim_and_waveform}. 

The stimulus cycle duration is determined by the presentation rate (here set to $20\,$Hz) and the length of the sequence, yielding $31/20 = 1.55$\,s for a cycle.
Usually, a presentation rate of $60$\,Hz is used, together with a $63$-bit code~\citep{martinez-cagigal2021}. However, in order to maintain a zooming motion, we reduced the presentation rate, such that enough frames could be used to show the motion, and to stay close to the frequency range used in SSMVEP literature. 

\subsubsection{Frequency Stimulation for SSMVEP and SSVEP}
In frequency-based BCI systems, such as SSVEP and SSMVEP, visual stimuli are modulated either through flicker or continuous motion according to constant, predefined frequencies. 
In the case of SSMVEP, it is most common to use a motion that includes a motion reversal, such as zooming in and zooming out. To achieve a perceptually smooth motion, sine-based stimulation waves are commonly used. Nevertheless, the use of square- or triangle-based stimulation waves has also been studied~\citep{yan2020, han2022, kwon2022}. Here, the SSMVEP stimulation is done with a sine-wave, using the following:
\begin{equation}\label{eq:sineapprox}
    S(f, i) = \dfrac{1}{2} (1 + sin(2\pi f (i/V_{RR})))
\end{equation}
where $f$ is the stimulation frequency (Hz), $i$ the frame index, and $V_{RR}$ the screen's vertical refresh rate. This is based on \citet{nakanishi2014, chen2019} in which the approximation method for SSVEP is used, where a frequency is approximated with a variable number of frames in a period. We opted to use this strategy since it increases the number of frequencies that can be used with a specific refresh rate. 

Similar to c-MVEP, we did not allow motion in this full range for SSMVEP. Again, the stimulus size was limited to a minimum of $50\,\%$ and a maximum of $100\,\%$ of the original stimulus size.
For the offline study, a motion-inversion frequency of $10$\,Hz is chosen, specifically, $f = 5$\,Hz. For the online study, with four targets, motion inversion frequencies, or flip-frequencies, of $9.0, 10.0, 11.25, 12.0$\,Hz are used. 

For SSVEP, perceiving such flickering stimuli evokes a response with distinct oscillatory components at the fundamental flicker frequency and its harmonics~\citep{vialatte2010}. SSVEP stimulation in the range of $6$ to $15$\,Hz was shown to evoke the strongest responses~\citep{ajami2018, nakanishi2014a, stawicki2015, stawicki2019}. Here, to be close to the c-VEP, a stimulation of $f = 10$\,Hz was chosen for the offline analysis, and $9.0, 10.0, 11.25, 12.0$\,Hz for the online study. The stimulation was done using a square-wave, and without phase differences.

\subsection{Experimental Protocol}
This study included two experiments, where both experiments followed a within-subject design. First, we did an offline experiment, where for each of the stimulus protocols, a single stimulus was presented on the screen to analyse the characteristics of the responses, see \autoref{fig:exp_protocol}. The second experiment was an online BCI selection application, where for each of the stimulus protocols, four targets were presented simultaneously on the screen to evaluate the decoding performance, see \autoref{fig:exp_protocol}.

\subsubsection{Experiment 1: Offline Response Analysis}
The offline experiment was done to characterize the neural responses of the novel c-MVEP paradigm prior to the online BCI study. The data from this experiment were treated as exploratory and were evaluated qualitatively as descriptive physiology, without statistical comparisons. 

In the offline experiment, only one stimulus was presented in the center of the screen. This stimulus had a visual angle of $4.34^{\circ} \times 4.34^{\circ}$, at a viewing distance of 60\,cm. We chose this design to obtain clean signals while avoiding potential interference from neighboring stimuli across the different experimental protocols.
Subjects were asked to focus on this stimulus after a $1.5$\,s cueing period indicating the start of stimulation. After the green cue, the stimulus was reset to the original state (maximum size and black) for $1.0$\,s before the stimulation started. 

For the c-MVEP and c-VEP conditions, stimulation lasted three times the cycle length, here $3 \times (31/20) = 4.65$\,s.
For SSMVEP and SSVEP, stimulation was $5.0$\,s. For all conditions, there was an inter-trial interval (ITI) of $0.5$\,s, after which the next trial started. 

To avoid order effects, the order of the conditions was randomized for the first run, and presented in the reversed order of the first run in the second run. The experiment was designed as follows: for each condition, 10 trials were presented, after which a questionnaire for this condition was filled in. Then, the next condition was started, where again 10 consecutive trials were presented, followed by the questionnaire. After all conditions had been observed and the final questions were answered, all conditions were presented again, in reversed order, each containing 10 trials, but did not include another questionnaire. This resulted in 20 trials per condition for each participant, see \autoref{fig:exp_protocol}.

Prior to starting the experiment, participants filled in a pre-experiment questionnaire, which asked about their previous BCI experience, current tiredness level, and the amount of sleep in the previous night. These data were collected for internal documentation only and were not included in the subsequent analyses.
As mentioned above, after the first run of each of the conditions, the participants filled in a questionnaire to obtain the subjective rating. The questionnaire consisted of six questions that were answered on a 6-point Likert scale. 
The questions asked about the perceived visual discomfort caused by the stimulus, how easy it was to keep concentrating on the target, how disturbing the stimulation was, how often they experienced a loss of focus, how much they would like the stimulation in a BCI speller, and lastly, their overall rating of the stimulus. 
The questions and scale values are represented in the Supplementary Material.

\subsubsection{Experiment 2: Online BCI Application}
The online experiment explored the feasibility of using c-MVEP for BCI applications. All four stimulation paradigms were implemented to compare the newly proposed c-MVEP against the well-established protocols. In the experiments, the subjects were instructed to perform a copy-selection task. The numerical BCI system consisted of $N = 4$ stimuli with the same size as in the offline study, placed on a horizontal line on the screen with a spacing of $7.26^{\circ}$ between stimuli. To avoid order effects, the order of conditions was randomized over participants. All stimulus conditions used the stimulation as described in \autoref{stimGen}.

Prior to starting the experiment, the participants completed a pre-questionnaire that asked about their demographic information, previous BCI experience, level of tiredness, and hours slept the previous night.

All conditions used the same classification approach, a template-matching canonical correlation analysis (CCA) classifier. The templates were built from user-specific data. To calibrate the classifier, the experiment started with a training block specific to the condition, to collect labeled training data. In this training block, the four stimuli were presented on the screen. Participants were instructed to focus on the cued stimulus until flickering or moving stopped. The cueing was done in green and lasted for $1.5$\,s, after which the stimulus was restored to its original color and size for $1.0$\,s, before flickering or moving started.
For c-MVEP and c-VEP the stimulation lasted $3 \times (31/20) = 4.65$\,s, for SSMVEP and SSVEP, the stimulation lasted $5.0$\,s. In total, each stimulus was attended to six times, for a total of 24 training trials per condition. After the training block of a condition, the participants were asked to fill out a questionnaire to rate their comfort on that condition.

\begin{figure}
\centering
\includegraphics[width=\textwidth]{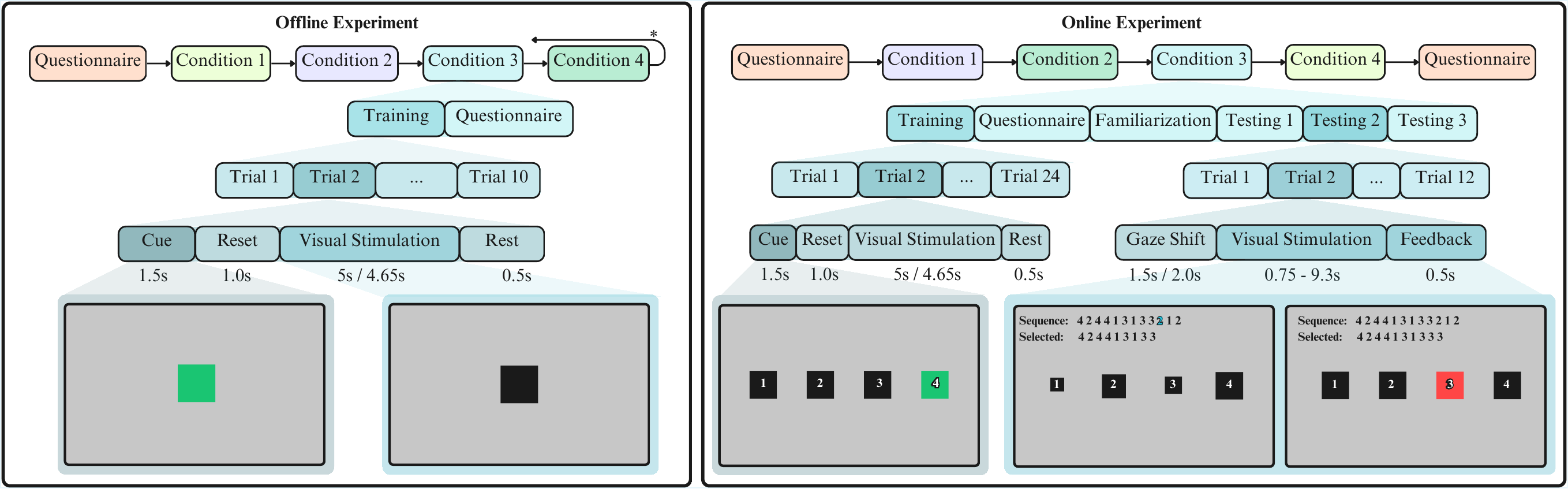}
\caption{\textbf{Experimental Protocol.} A graphical representation of the experimental protocol of the offline (left) and online (right) experiments. In both experiments, all four conditions, c-MVEP, c-VEP, SSMVEP, and SSVEP were presented randomly. The visual stimulation duration was dependent on the stimulation paradigm, for c-MVEP and c-VEP this was 4.65\,s, and 5\,s for SSMVEP and SSVEP. The left figure shows the offline protocol, where one stimulus was shown on the screen. The left screen shows the cueing in green. The right shows the stimulation. Depending on the stimulation condition, this was either zooming (for c-MVEP and SSMVEP) or black-white flickering (c-VEP and SSVEP). Please note that, in the offline study, after the four conditions were presented in random order, they were presented again in reverse order, without the questionnaire. The figure on the right shows the online protocol, where four stimuli were presented simultaneously. From left to right, the screens depict the cueing in green for the training block, stimulation in the testing block, and the feedback for an incorrect selection.} \label{fig:exp_protocol}
\end{figure}

After the questionnaire, the participants were introduced to the BCI with a 4-digit familiarization sequence to ensure they understood the selection procedure. This was followed by the start of the testing block, which consisted of three runs, each with a 12-digit-long sequence. In each run, the full sequence to be entered was displayed at the top of the screen. To facilitate recognition of the next selection, the digit to be selected was highlighted in bold and blue within the sequence during gaze shifts and stimulation, see~\autoref{fig:exp_protocol}. The subjects had to shift their gaze to the corresponding target during a gaze-shifting period of $1.5$\,s. 

Following the gaze-shift period, all $N=4$ stimuli started flickering or moving simultaneously. When a classification was reached, the stimulation of all targets stopped, and the prediction was indicated using visual and auditory feedback and printed on the screen. A correct prediction produced a positive sound, and the stimulus was indicated in green for $0.5$\,s. For an incorrect prediction, the predicted target was indicated in red, together with a neural sound to provide clear auditory feedback without inducing negative reinforcement or user frustration. Incorrect selections could not be corrected, so participants had to continue to the next target. After a full sequence was selected, the participant could continue to the next sequence at their own pace, see \autoref{fig:exp_protocol}.
Between sequences, three participants explicitly reported that the gaze-shifting period was too short. For those participants, the gaze-shift was extended to $2.0$\,s, to ensure they could comfortably continue the experiment.

After each training block for a condition, participants completed a questionnaire to provide their subjective rating. This ensured that the condition's BCI performance did not influence the answers. 
A total of seven questions were answered on a 6-point Likert scale. The questions inquired after their level of fatigue, how comfortable the stimulus was to look at, how easy it was to keep concentrating on the target, how disturbing the stimulation was, how often they experienced a loss of focus, how much they would like the stimulation in a BCI speller, and lastly, their general rating of the stimulus. 

After the experiment, the participants filled in a post-experiment questionnaire. This inquired about the differences between zooming and flickering, specifically regarding which was easier to keep focus on, which caused the most eye strain, and what participants preferred. Lastly, they were also asked whether they observed a difference between the two flickering conditions and the two zooming conditions. The questionnaire and post-questionnaire with the question and answers scales are presented in the Supplementary Material.

\subsection{Data Processing and System Evaluation}
\subsubsection{Experiment 1: Offline Response Analysis}
The data obtained in the offline experiment were examined in the time, frequency, and spatial domains. 
In the time-domain, we calculated the evoked waveforms at channel Oz to get an insight in the response to one stimulation cycle. A stimulation cycle was $31/20 = 1.55$\,s for c-MVEP and c-VEP, and we used $1.0$\,s for SSMVEP and SSVEP. All trials were cut into non-overlapping segments of these cycles $\bm{X}_i \in \mathbb{R}^{C \times T}$, with $C$ channels and $T$ samples, resulting in $60$ segments for c-MVEP and c-VEP and $100$ segments for SSMVEP and SSVEP.
Prior to computing the grand average for SSMVEP and SSVEP, each subject's waveform was temporally shifted by up to $\pm50$\,ms. This shift maximized the cross-correlation of the P1 peak with the group average, effectively correcting for inter-subject phase and latency differences.

To examine the frequency spectrum, the signal-to-noise ratio (SNR) at Oz was calculated for each subject and condition. The power spectral density (PSD) for each trial was estimated using a Hamming-windowed Fast Fourier Transform, zero-padded to 10 seconds for a uniform 0.1\,Hz frequency resolution. The resulting single-trial power spectra were averaged to reduce background noise and enhance the VEP spectral peaks. The SNR was calculated differently for the code-modulated and frequency-based paradigms to prevent spectral leakage, while allowing paradigm-specific spectral characteristics. For c-MVEP and c-VEP, we estimated the noise as the median power of 15 bins on each side, excluding the 2 adjacent bins. For SSMVEP and SSVEP, the noise was estimated from 20 neighbouring frequency bins on either side, excluding the 4 immediately adjacent bins to avoid signal leakage.  

To investigate the spatial distribution of activation across the paradigms, we created topographic maps for the SNR. For each condition, the spatial analysis was limited to the primary target frequency driving the visual response, $20$\,Hz for c-MVEP and c-VEP, $5$\,Hz for SSMVEP, and $10$\,Hz for SSVEP.

\subsubsection{Experiment 2: Online BCI Application}
In c-VEP-, SSMVEP-, and SSVEP-based BCIs, canonical correlation analysis (CCA) has been successfully applied for template matching~\citep{nakanishi2015, bin2009} or frequency matching~\citep{lin2007, nakanishi2015}.
CCA is a multivariate statistical analysis method that reveals the underlying correlation between two sets of multidimensional variables~\citep{hotelling1936}. For two sets of signals $\mathbf{X}$ and $\mathbf{Y}$, the goal is to find linear projections $\mathbf{w}_x$ and $\mathbf{w}_y$ that maximise the correlation between $\mathbf{w}_{x}^{\top}\mathbf{X}$ and $\mathbf{w}_{y}^{\top}\mathbf{Y}$. Normally, $\mathbf{X}$ denotes the multi-channel EEG data, and $\mathbf{Y}$ the reference signals.

We adopted a unified decoding framework across all experimental conditions, using a template-based CCA approach. For the well-known c-VEP, SSVEP, and SSMVEP conditions, there exist methods specifically tailored to the paradigm that reach state-of-the-art performances, however, here we opted to use the same approach for each condition. This choice was motivated by the absence of evidence that any of the state-of-the-art methods from the c-VEP domain, such as reconvolution-based CCA~\citep{thielen2015,thielen2021}, generalizes effectively to the c-MVEP condition. Moreover, using a consistent decoding method across the conditions ensured that observed performance differences reflect the stimulation paradigm rather than the decoder.

Specifically, the CCA was implemented as a template-matching classifier, where it builds the templates as the averaged response to repeated stimulation of the same target stimulus: 
\begin{equation}
    \bm{T}_i = \dfrac{1}{R} \sum_r \bm{X}_r,
\end{equation}
where $\bm{T}_i \in \mathbb{R}^{C \times M}$ and $\bm{X}_r \in \mathbb{R}^{C \times M}$ with $r = 1, \dots, R$ the number of single trials for the $i$-th class. Here, $C$ denotes the number of channels, and $M$ the number of samples.

Then, spatial filters are learned using CCA for the raw data in $\bm{X}$ and templates of $\bm{T}$. First, the templates $\bm{T}_i$ are stacked and repeated, according to the order of all $j = 1, \dots, J$ training trial in $\bm{X}$:
\begin{equation}
    \bm{T} = [\bm{T}_{y_1}, \bm{T}_{y_2}, \dots, \bm{T}_{y_J}],
\end{equation}
where $\bm{T} \in \mathbb{R}^{C \times M \cdot J}$, and $\bm{T}_{y_j}$ is the template of the class that is specified by the label $y_j$ of the $j$th trial. All trials in $\bm{X}$ are concatenated to $\bm{S} \in \mathbb{R}^{C \times M \cdot J}$.
Then, using $\bm{T}$ and $\bm{S}$, an optimized spatial filter $\bm{w}$ can be found using CCA:
\begin{equation}
    \max_{\bm{w }, \bm{v}} \text{corr} (\bm{w}^{\top} \bm{S}, \bm{v}^{\top} \bm{T}) 
    = \max_{\bm{w }, \bm{v}} \dfrac{\bm{w}^{\top} \bm{S}\bm{T}^{\top} \bm{v}}{\sqrt{\bm{w}^{\top} \bm{S}\bm{S}^{\top} \bm{w} \cdot \bm{v}^{\top} \bm{T}\bm{T}^{\top} \bm{v}}}.
\end{equation}

During the online classification, restricting to the first CCA component only, spatial filters $\bm{w} \in \mathbb{R}^{C}$ and $\bm{v} \in \mathbb{R}^{C}$ can be used to spatially filter a single trial  $\bm{X} \in \mathbb{R}^{C \times M}$ and multi-channel templates $\bm{T}_i$, respectively: 
\begin{equation}
    \bm{x} = \bm{w}^{\top}\bm{X}, \qquad  \bm{t}_i = \bm{v}^{\top}\bm{T}_i,
\end{equation}
where $\bm{x} \in \mathbb{R}^{M}$ represents the weighted sum of the channels, and $\bm{t}_i \in \mathbb{R}^{M}$ the spatially filtered template response of the $i$th class. 
Then, the single trial is classified by selecting the template response $\bm{t}_i$ that maximizes Pearson's correlation:
\begin{equation}
    \hat{y} 
    = \arg \max_{i} \rho_i
    = \arg \max_{i} \dfrac{\bm{x}^{\top}\bm{t}_i}{\sqrt{\bm{x}^{\top}\bm{x} \cdot \bm{t}_{i}^{\top}\bm{t}_{i}}}.
\end{equation}

Furthermore, we applied a dynamic stopping approach, similar to the margin method used in~\citet{thielen2015}. Specifically, a classification result from the CCA was only accepted if a certain confidence threshold was met. A trial was accepted, and feedback was presented when the margin $\rho_m - \rho_r$, that is the difference between the maximum correlation $\rho_m$ and runner-up $\rho_r$, was larger than or equal to $0.3$. This threshold was set given pilot data, and was adjusted to $0.2$ if, during the familiarisation sequence, it was too strict, leading to trial-time outs. The threshold was tested for every $0.05$\,s into the trial. If the end of a trial was reached and the threshold was not exceeded, no decision was forced.  

The minimum window for classification was set to $0.75$\,s, the maximum time for classification was set to $9.3$\,s, which equals to six m-sequence cycles. When this maximum time was reached, and the confidence did not exceed the threshold, the classifier returned a no-decision, indicated by a `0', and the participant would continue with the next digit.

Furthermore, a combined expanding and sliding window approach was implemented. 
Initially, the window length was set to $0.75$\,s, after which the classifier attempted to decode the data. If the confidence threshold was not satisfied, the window was incrementally extended in steps of $0.05$\,s (equivalent to $30$ samples). At each increment, the classifier re-evaluated the enlarged window, and this expansion continued until the threshold was met. 
If the window reached a maximum length of $4.65$\,s for c-MVEP and c-VEP, or $5.0$\,s for SSMVEP and SSVEP, without satisfying the confidence threshold, it was assumed that the initial portion of the window did not contain sufficiently informative data. Therefore, a sliding window approach was then used. The window was shifted forward in steps of $0.05$\,s, while maintaining the maximum window length (first-in-first-out). This process continued iteratively until the threshold was satisfied or the total maximum classification time of $9.3$\,s was reached.

In addition to accuracy, the online target selection performance was evaluated with the (practical) information transfer rate (ITR)~\citep{wolpaw2002}, measured in bits per minute,
\begin{equation}
    ITR = \dfrac{\log_2 N + p \log_2 p + (1-p)\log_2 (\tfrac{1-p}{N-1})}{t/60},
\end{equation}
where $N$ denotes the number of stimulus classes, $\mathit{p}$ denotes the classification accuracy, and $\mathit{t}$ denotes the average selection time (in seconds), including the inter-trial time (ITI) of $2.0$\,s to $2.5$\,s. For more information on ITR and a publicly available tool of our lab, see our website~\footnote{\url{https://bci-lab.hochschule-rhein-waal.de/en/itr.html}}. 

All statistical analyses were performed in Python using the SciPy and statsmodels libraries. First, Shapiro-Wilk tests were conducted to assess the normality of the data distributions for accuracy, ITR, and average selection time. As the data violated the assumption of normality, non-parametric tests were used for all analyses. That is, a Friedman test was used as an omnibus test to detect significant differences among the four conditions. Following the significant Friedman results, post-hoc pairwise comparisons were computed using Wilcoxon signed-rank tests. To correct for multiple comparisons, p-values were adjusted using the Bonferroni correction (significance threshold $\alpha_{adj} = .05/6 \approx .0083$).

\section{Results}
\subsection{Experiment 1: Offline Response Analysis}
\subsubsection{Characteristics of Response}
\begin{figure}[t]
\centering
\renewcommand\tabularxcolumn[1]{m{#1}}
\setlength{\tabcolsep}{3pt} 
\begin{tabularx}{\linewidth}{
    @{}
    >{\centering\arraybackslash}m{0.02\linewidth}
    *{3}{>{\centering\arraybackslash}X} 
    @{}
}
    & \textbf{A)} Evoked Waveform (Oz)
    & \textbf{B)} SNR Spectrum (Oz)
    &\textbf{C)} SNR Spatial Distribution \\[1ex] 
    \rotatebox[origin=c]{90}{\textsc{c-MVEP}} & 
    \includegraphics[width=\linewidth, valign=m]{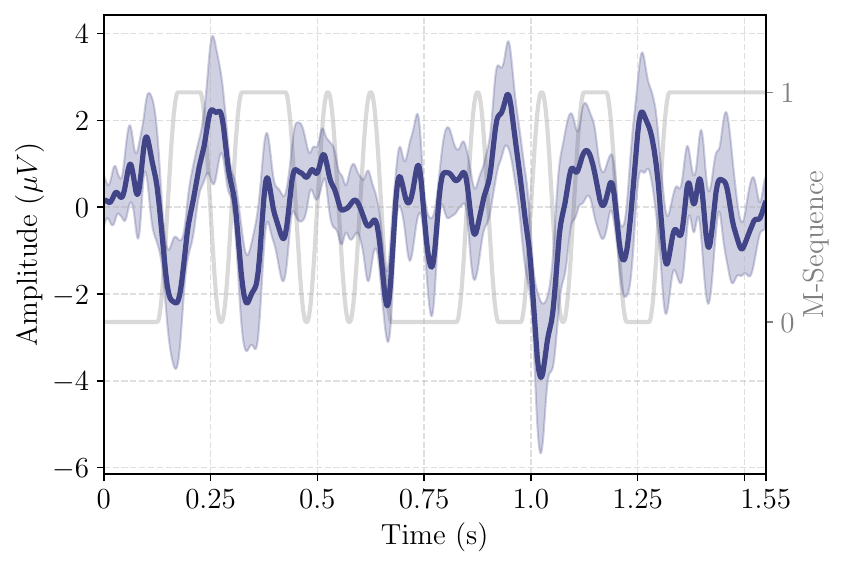} &
    \includegraphics[width=\linewidth, valign=m]{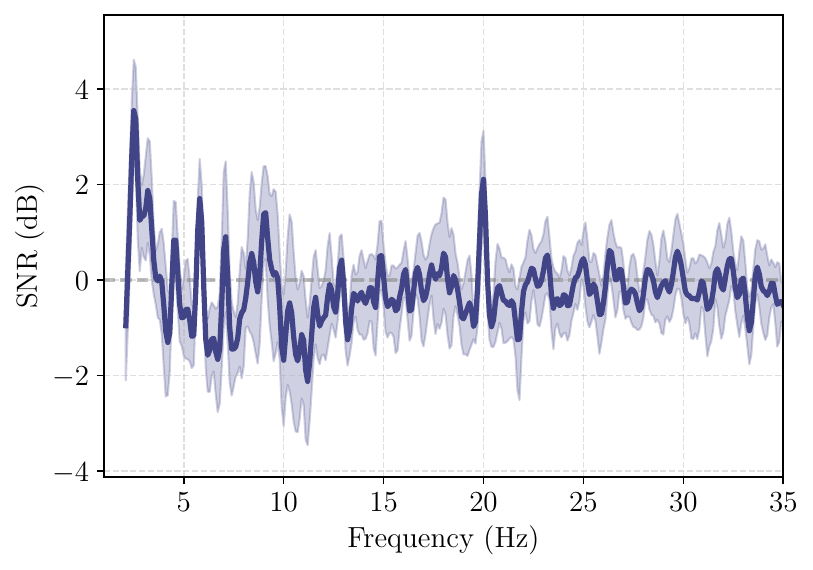} &
    \includegraphics[width=0.75\linewidth, valign=m]{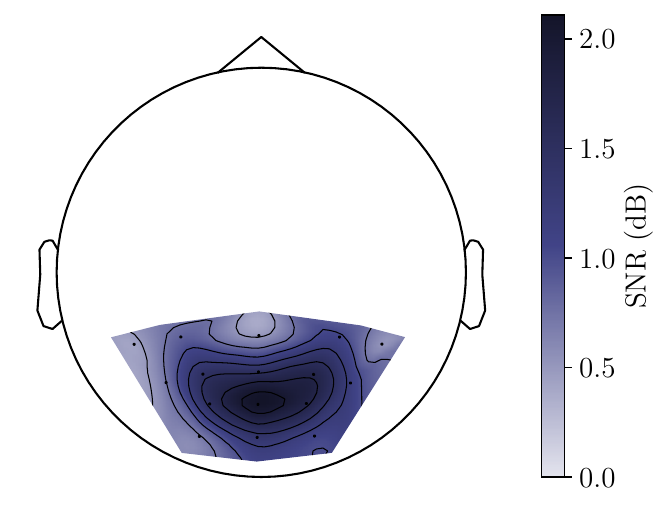}\\

    \rotatebox[origin=c]{90}{\textsc{c-VEP}} & 
    \includegraphics[width=\linewidth, valign=m]{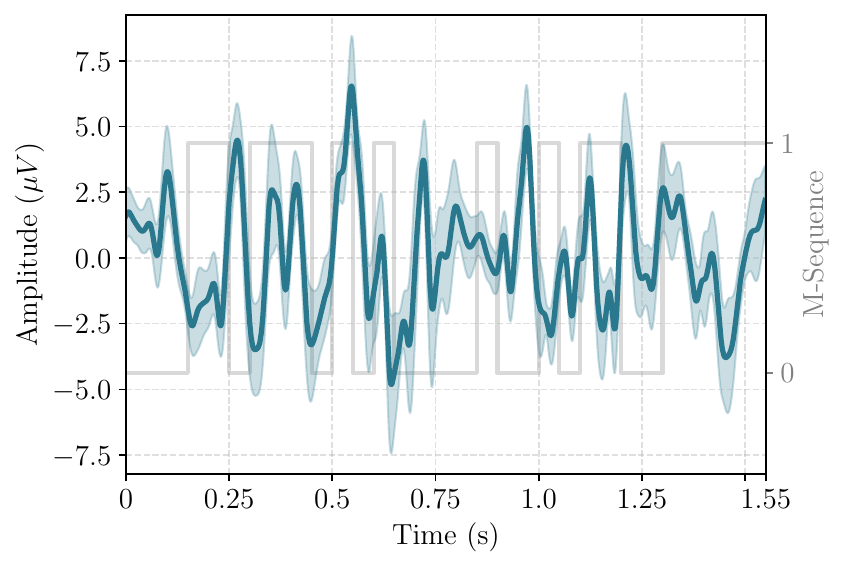} &
    \includegraphics[width=\linewidth, valign=m]{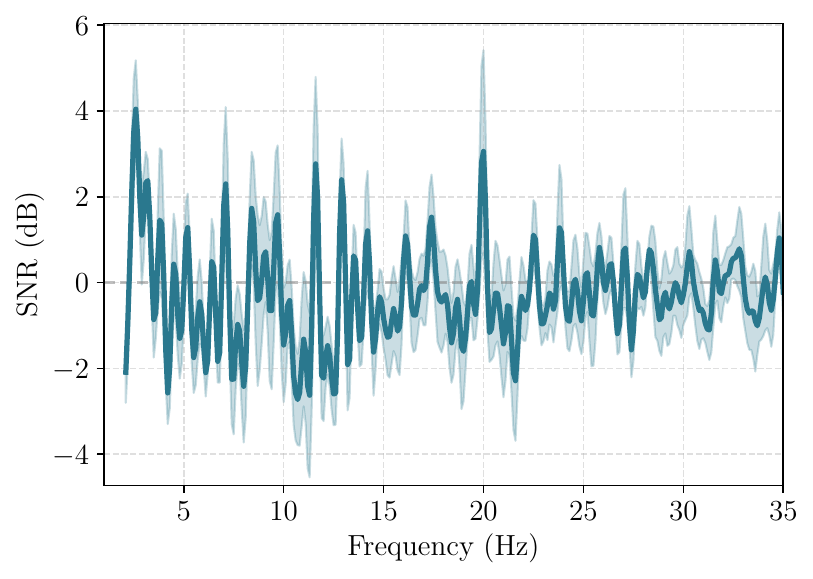} &
    \includegraphics[width=0.75\linewidth, valign=m]{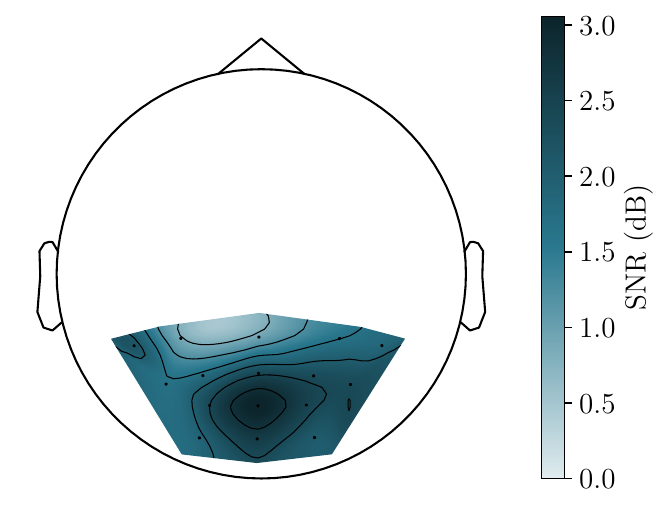} \\

\end{tabularx}
\caption{\textbf{Signal characteristics for c-MVEP and c-VEP}. The thick lines show the mean response, the shaded area the 95\,\% confidence intervals. Columns show (A) the grand-average evoked waveform at channel Oz for a full cycle of $1.55$\,s. Here, the smoothened m-sequence (c-MVEP) and the binary m-sequence (c-VEP) are plotted in grey in the background. Column (B) shows the SNR spectrum at channel Oz. Column (C) shows the spatial distribution of the $20$\,Hz power.}
\label{fig:cvep_characteristics}
\end{figure}

\begin{figure}[t]
\centering
\renewcommand\tabularxcolumn[1]{m{#1}}
\setlength{\tabcolsep}{3pt} 
\begin{tabularx}{\linewidth}{
    @{}
    >{\centering\arraybackslash}m{0.02\linewidth}
    *{3}{>{\centering\arraybackslash}X} 
    @{}
}
    & \textbf{A)} Evoked Waveform (Oz)
    & \textbf{B)} SNR Spectrum (Oz)
    &\textbf{C)} SNR Spatial Distribution \\[1ex] 
    \rotatebox[origin=c]{90}{\textsc{SSMVEP}} & 
    \includegraphics[width=\linewidth, valign=m]{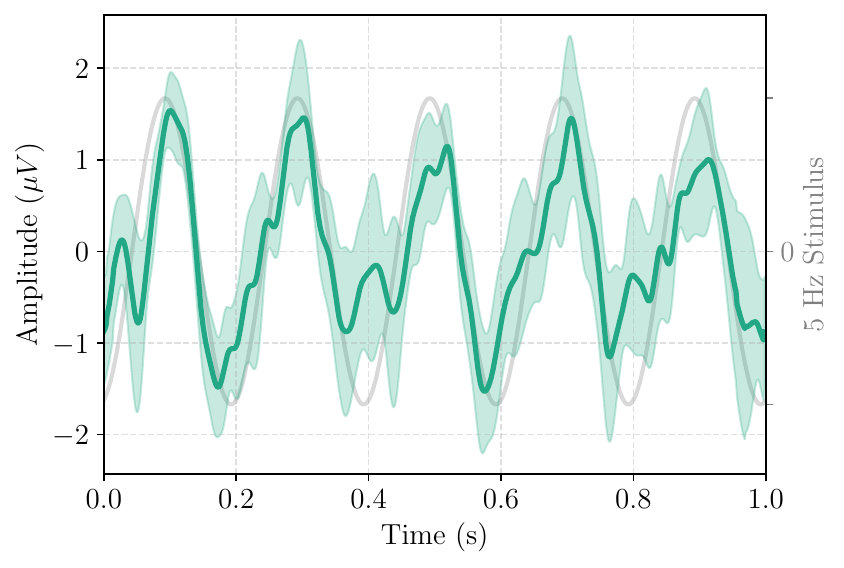} &
    \includegraphics[width=\linewidth, valign=m]{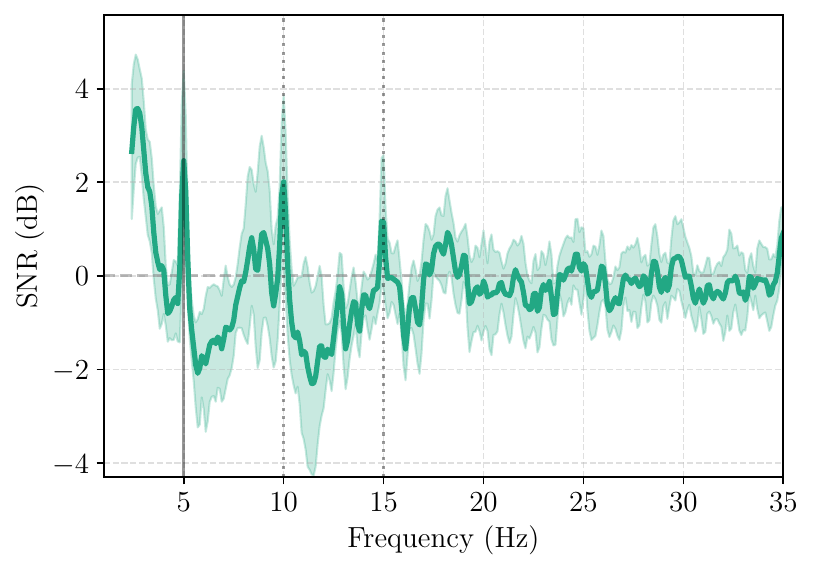} &
    \includegraphics[width=0.75\linewidth, valign=m]{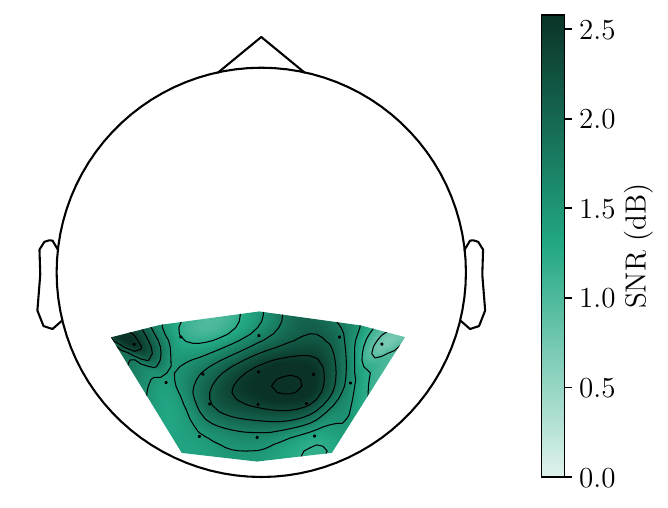}\\

    \rotatebox[origin=c]{90}{\textsc{SSVEP}} & 
    \includegraphics[width=\linewidth, valign=m]{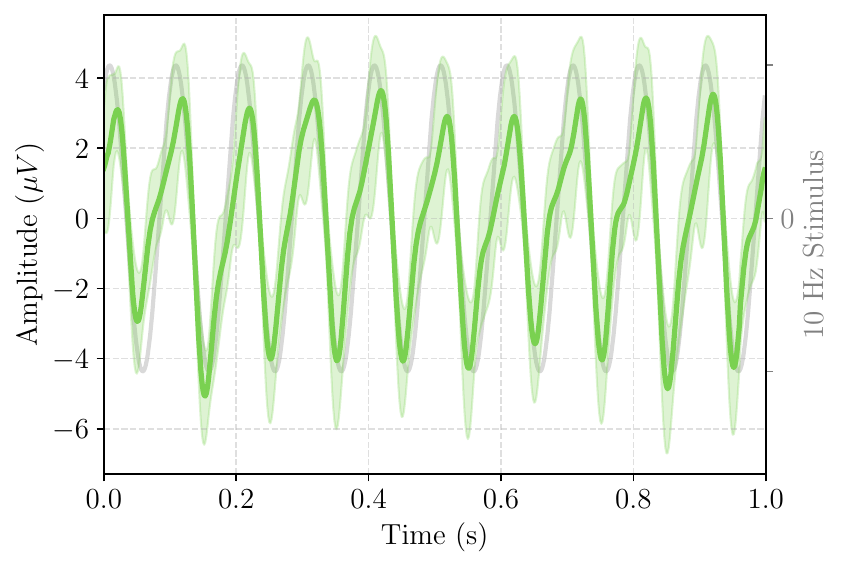} &
    \includegraphics[width=\linewidth, valign=m]{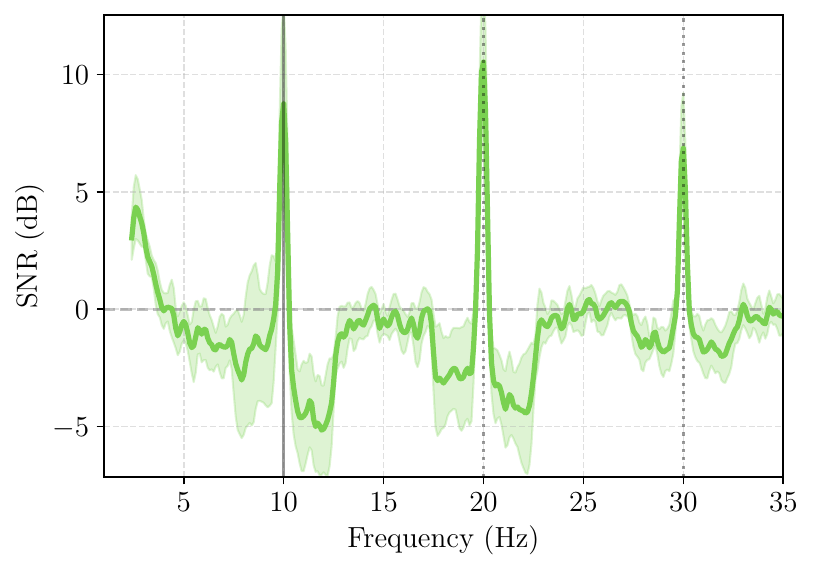} &
    \includegraphics[width=0.75\linewidth, valign=m]{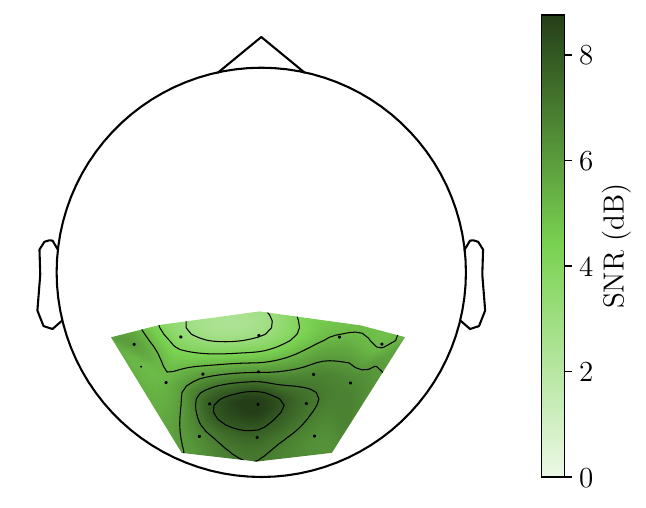} \\

\end{tabularx}
\caption{\textbf{Signal characteristics for SSMVEP and SSVEP}. The thick lines show the mean response, the shaded area the 95\,\% confidence intervals. Columns show (A) the grand-average evoked waveform at channel Oz for $1.0$\,s, with the original stimulation waveform in light gray. Column (B) shows the SNR spectrum at channel Oz. Column (C) shows the spatial distribution of the $5$\,Hz (SSMVEP) and $10$\,Hz (SSVEP) power.}
\label{fig:ssvep_characteristics}
\end{figure}

The response characteristics of the c-MVEP and c-VEP are shown in \autoref{fig:cvep_characteristics}. The signal response characteristics SSMVEP and the SSVEP conditions are shown in \autoref{fig:ssvep_characteristics}.  Please refer to the Supplementary Material for figures of the PSD spectrum and PSD spatial distribution.
For all conditions, the grand-average evoked waveform at channel Oz, the SNR spectra at channel Oz, and spatial distributions were investigated.
The following observations are presented as a descriptive physiological baseline to verify the visual evoked responses.

The evoked waveform was a grand-average over all segments, trials, and subjects. For c-MVEP and c-VEP, each trial was cut into full m-sequence cycles, so $31/20 = 1.55$\,s. For SSMVEP and SSVEP, each trial was cut into $1.0$\,s segments.

The evoked waveforms for c-MVEP and c-VEP did not show any frequency-specific oscillations, which was expected due to their code-modulated, non-periodic stimulation structure. This was supported by the SNR and PSD spectra, in which no specific narrow-band peaks can be identified, but rather the energy was observed to be distributed more broadly across the frequency bands. The evoked waveform of c-MVEP presents a larger negative peak around $1.0$\,s, which did not occur in c-VEP. Its preceding positive peak did appear to be visible in both. The remaining waveform presents a similar pattern, although with clear, distinct differences.

The frequency spectra for c-MVEP and c-VEP showed a broad-band response to the m-sequence. For c-MVEP, it could be observed that the broad-band response was more focused at the lower frequencies, whereas for c-VEP it also spread more beyond $10$\,Hz. In both, a peak at the $20$\,Hz presentation rate was observed, with descriptive mean values of $2.11\pm0.52$\,dB for c-MVEP and $3.06\pm1.20$\,dB for c-VEP.
Lastly, we inspected the spatial distributions of the $20$\,Hz activation. The c-MVEP activation appeared distributed across multiple electrodes, with the main activation at Oz. The c-VEP activation, Oz contained the majority of the $20$\,Hz activation, with visibly less spread to neighboring electrodes.

The evoked waveforms for SSMVEP and SSVEP, aligned across participants prior to averaging, were characterized by sinusoidal patterns. For SSVEP, the waveform predominantly reflected the stimulus's fundamental frequency, whereas SSMVEP presented a more complex structure. This difference may partly be caused by an overall lower SNR observed for SSMVEP compared with SSVEP (descriptive mean values of SNR of $2.46 \pm 0.98$\,dB and mean peak-to-peak $5.55 \pm 0.41\mu V$ for SSMVEP, and SNR of $8.75 \pm 2.14$\,dB and mean peak-to-peak $12.23\pm1.14\mu V$ for SSVEP).

From the frequency spectrum of SSMVEP and SSVEP, frequency responses at the fundamental frequency and its harmonics, can be observed. 
A descriptive evaluation showed the SSMVEP response was characterized by the distinct harmonics at $10$\,Hz (SNR $2.01\pm0.94$\,dB) and at $15$\,Hz (SNR $1.15\pm0.73$\,dB). For SSVEP, the harmonics appeared stronger and more distinct at $20$\,Hz (SNR $10.54\pm1.81$\,dB) and $30$\,Hz (SNR $6.86\pm1.20$\,dB).

\subsubsection{Subjective Rating}

\begin{figure}[h]
\centering
\includegraphics[width=\textwidth]{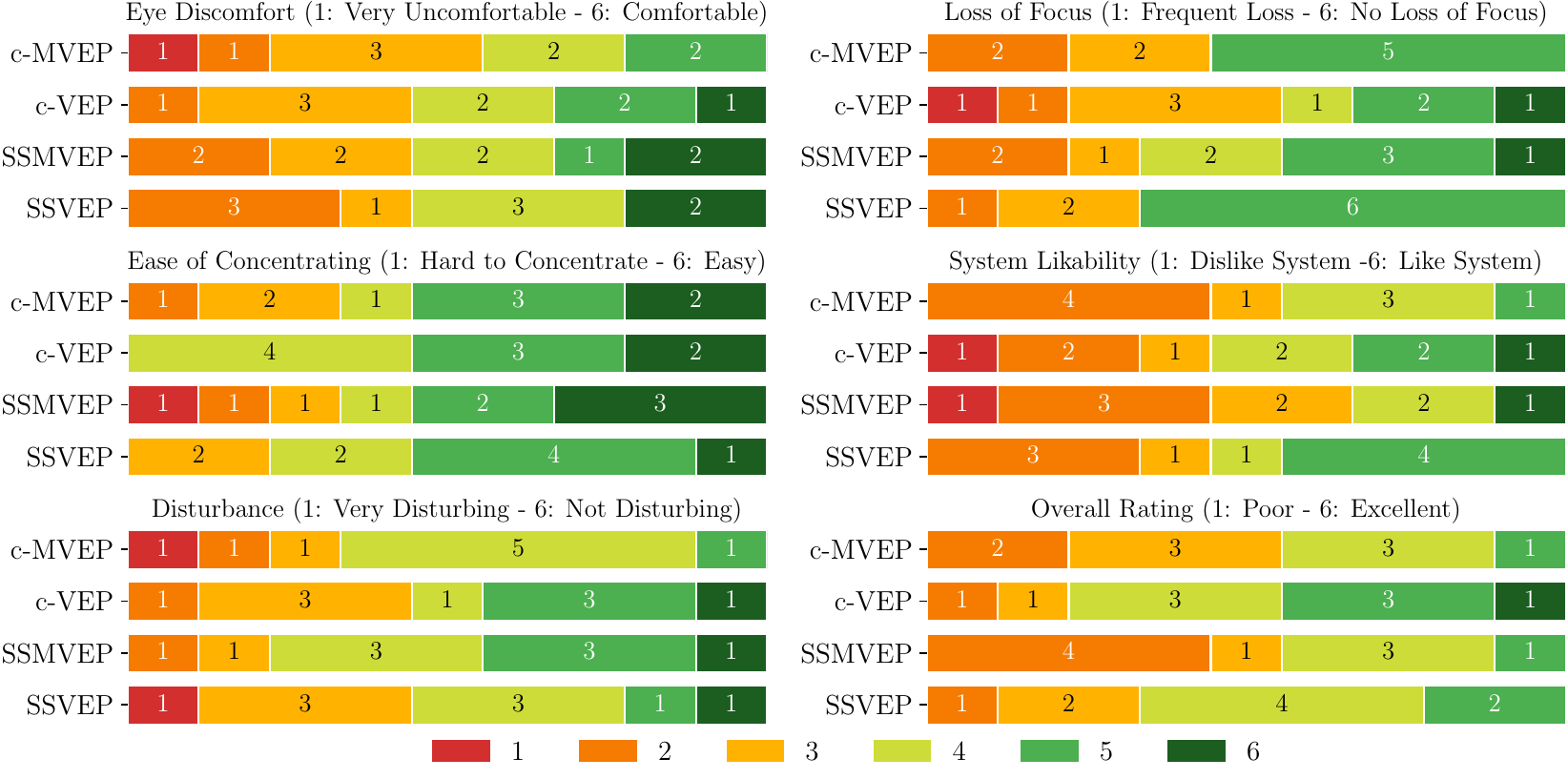}
\caption{\textbf{Results Offline Questionnaire.} The answers to the questionnaire from the offline experiment. Questions were answered on a 6-point Likert scale. Here, 1 indicates a more negative rating, and 6 a more positive rating. The numbers in the bars indicate the number of participants giving that specific score.}\label{fig:offline_questionnaire}
\end{figure}

After the first run of a condition in the offline experiment, the participants were presented with a questionnaire that assessed their subjective rating of the stimulation. For each condition, six questions inquired about the visual comfort of the stimulus, ease of concentrating, disturbance, perceiving loss of focus, preference to use the system with more targets, and an overall score. Each question was answered on a 6-point Likert scale. \autoref{fig:offline_questionnaire} shows the results of the nine subjects, the average and standard deviation values can be found in \autoref{tab:questionnaireOffline}. 

To ensure directional consistency across all metrics, the responses to the `Loss of Focus' question were reverse-scored prior to analysis, such that a higher score indicated more favorable ratings for all measures.

Although the offline experiment was primarily exploratory and constrained by a small sample size (N = 9), a statistical analysis was conducted. Given the limited sample size, the results should be interpreted with caution.
The non-parametric Friedman test was used to test for a significant effect of condition for each question. No significant differences were found between the conditions for eye-discomfort ($\chi^2 = 0.712, p=0.870$), ease of concentration on the target ($\chi^2 = 0.095, p=0.992$), disturbance caused by the flickering or motion ($\chi^2 = 2.542, p=0.468$), loss of focus ($\chi^2 = 2.000, p=0.572$),  system likability ($\chi^2 = 0.958, p=0.811$), or overall impression ($\chi^2 = 2.508, p=0.474$).

\begin{table}[]
\caption{\textbf{Questionnaire Results Offline} Average (SD) results of the offline questionnaire. Higher scores indicate more favourable ratings.}\label{tab:questionnaireOffline}
\begin{tabular}{lcccccc}
        \hline\\[-1.5ex]
       & \textbf{Discomfort}         & \textbf{Concentrating}   & \textbf{Disturbance}    & \textbf{Focus}           & \textbf{Likability}     & \textbf{Rating}           \\ \hline\\[-1.5ex]  
\textbf{c-MVEP} & $3.33$ $(1.25)$ & $4.33$ $(1.33)$ & $3.44$ $(1.17)$ & $3.89$ $(1.29)$ & $3.11$ $(1.10)$ & $3.33$ $(0.94)$   \\
\textbf{c-VEP}  & $3.89$ $(1.20)$ & $4.78$ $(0.79)$ & $4.00$ $(1.25)$ & $3.56$ $(1.50)$ & $3.56$ $(1.57)$ & $4.22$ $(1.13)$   \\
\textbf{SSMVEP} & $3.89$ $(1.45)$ & $4.22$ $(1.75)$ & $4.22$ $(1.13)$ & $4.00$ $(1.33)$ & $3.00$ $(1.41)$ & $3.11$ $(1.10)$  \\
\textbf{SSVEP}  & $3.67$ $(1.49)$ & $4.44$ $(0.96)$ & $3.67$ $(1.33)$ & $4.22$ $(1.13)$ & $3.67$ $(1.33)$ & $3.78$ $(0.92)$
\end{tabular}
\end{table}

\subsection{Experiment 2: Online BCI Application}
\subsubsection{BCI Performance}
\begin{figure}[t]
\centering
\renewcommand\tabularxcolumn[1]{m{#1}}
\setlength{\tabcolsep}{3pt} 
\begin{tabularx}{\linewidth}{
    @{}
    *{3}{>{\centering\arraybackslash}X} 
    @{}
}

    \textbf{A)} Accuracy
    & \textbf{B)} Selection Time
    & \textbf{C)} ITR \\[1ex] 
    \includegraphics[width=\linewidth, valign=m]{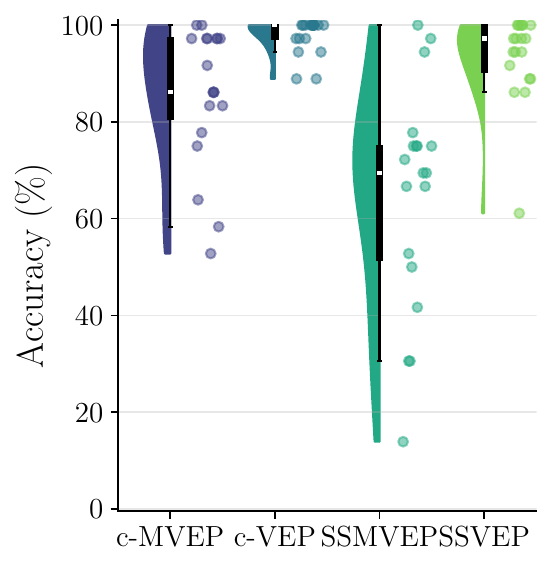} &
    \includegraphics[width=\linewidth, valign=m]{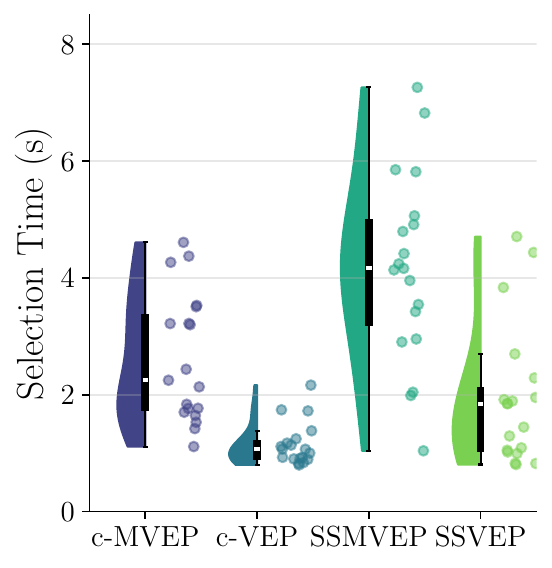} &
    \includegraphics[width=\linewidth, valign=m]{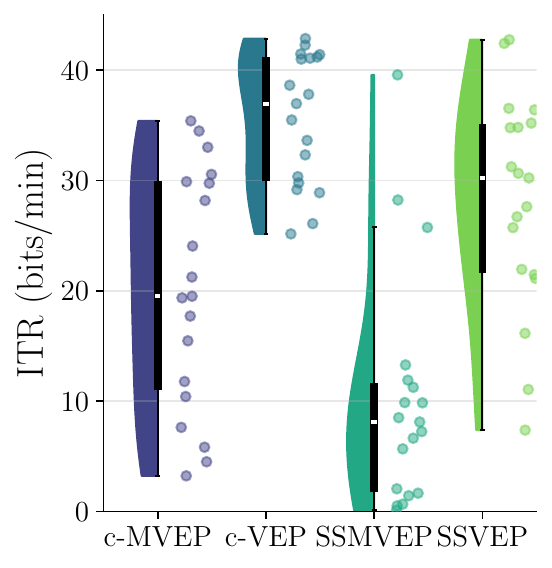}\\

\end{tabularx}
\caption{\textbf{Accuracy, Selection Time, and ITR Results for the Online Experiment.} Raincloud plots with the performances of the online BCI selection experiment for each condition (c-MVEP in dark-blue, c-VEP in teal, SSMVEP in green, SSVEP in lime green) per performance metric (Accuracy, Selection Time, ITR). Specifically, A) shows the distribution of the decoding accuracies for all participants and 12-digit sequences. Graph B) depicts the average selection time, excluding ITI time. Graph C) shows the average ITR, which included ITI time. The half-split violin plot represents the density distribution of the data, the standard boxplot indicates the first and third quartiles, with the white horizontal lines denoting the median, and the jittered scatter points represent the averaged metric for each participant.}\label{fig:performances_online_acc_selec}
\end{figure}

The online selection task was performed for all four conditions c-MVEP, c-VEP, SSMVEP, and SSVEP. Using the four targets on the screen, the participants had to select numbers 1 to 4 according to a randomised 12-digit-long sequence. They fixated on the corresponding target, evoking a specific VEP response. Using CCA, this response was decoded, and feedback was given to the participant. 
The results in the online selection tasks are presented in \autoref{fig:performances_online_acc_selec}. \autoref{tab:performance_table} denotes the mean performance and standard deviation for each condition and performance metric.

For the accuracy, \autoref{fig:performances_online_acc_selec}A), a Friedman test revealed a significant main effect of condition ($\chi^2 = 34.59$, $p < .001$). Mean accuracies were highest for c-VEP ($97.81 \pm 3.65\,\%$), followed by SSVEP ($93.42 \pm 9.18\,\%$), c-MVEP ($85.67 \pm 14.41\,\%$), and SSMVEP ($64.91 \pm 23.21\,\%$) (mean $\pm$ SD). 
Post-hoc pairwise comparisons were conducted using a one-sided Wilcoxon signed-rank test, with a Bonferroni correction ($\alpha_{adj} = .0083$). 
We found that c-MVEP reached significantly higher accuracies than SSMVEP ($p<.001$). SSVEP significantly outperformed SSMVEP ($p \leq .001$) and c-MVEP ($p = .002$). Additionally, c-VEP outperformed SSMVEP ($p \leq .001$) and c-MVEP ($p = .001$). Lastly, there was no significant difference between SSVEP and c-VEP ($p=.018$).

The selection time, see \autoref{fig:performances_online_acc_selec}B), was defined as the stimulation or decoding time needed for selection, so not including the ITI. The Friedman test showed significant differences between conditions ($\chi^2 = 38.56$, $p < .001$).
c-VEP had the shortest mean selection times ($1.15 \pm 0.37$\,s), followed by SSVEP ($1.94 \pm 1.20$\,s), c-MVEP ($2.61 \pm 1.09$\,s), and SSMVEP ($4.18 \pm 1.62$\,s) (mean $\pm$ SD). Post-hoc Wilcoxon signed-rank tests with Bonferroni correction revealed that c-MVEP reached significantly faster selection times than SSMVEP ($p<.001$), but was outperformed by SSVEP ($p=.027$). SSVEP furthermore reached significantly faster selection times than SSMVEP ($p<.001$). Lastly, c-VEP significantly outperformed c-MVEP, SSMVEP, and SSVEP (all $p\leq .001$).

The ITR results, which included the ITI, see \autoref{fig:performances_online_acc_selec}C), showed a significant difference between conditions ($\chi^2 = 41.53$, $p < .001$). Average ITR values were highest for c-VEP ($35.56 \pm 5.93$\,bits/min), followed by SSVEP ($28.12 \pm 9.78$\,bits/min), c-MVEP ($20.11 \pm 10.71$\,bits/min), and SSMVEP ($10.13 \pm 10.53$\,bits/min) (mean $\pm$ SD). 
The post-hoc Wilcoxon signed-rank tests with Bonferroni correction revealed that c-MVEP significantly outperformed SSMVEP ($p<.001$), but was significantly outperformed by c-VEP and SSVEP ($p<.001$). Both SSVEP and c-VEP also reached significantly higher ITR than SSMVEP ($p<.001$), and c-VEP significantly outperformed SSVEP ($p<.001$).

To assess whether the effect of adding motion differed between SSVEP and c-VEP paradigms, a subject-wise motion effect score (motion - flicker) was computed for each paradigm. These difference scores were compared using a two-sided Wilcoxon signed-rank test.

Adding motion reduced accuracy by $28.51 \pm 24.04\,\%$ in SSVEP paradigm, compared to $12.13 \pm 15.08\,\%$ in the c-VEP paradigm (mean $\pm$ SD). The reduction in accuracy due to adding motion was significantly different between the paradigms ($p = 0.031$).
Selection time increased by $2.24 \pm 1.86$\,s in the SSVEP paradigm when motion was added, compared to $1.46 \pm 1.06$\,s in the c-VEP paradigm; however, this difference was not statistically significant ($p = .060$).
Similarly, the ITR decreased by $17.99 \pm 10.83$\,bits/min in the SSVEP paradigm and by $15.45 \pm 10.57$\,bits/min in the c-VEP paradigm, with no significant difference in motion-related reduction between the paradigms ($p=.623$).

During the online task, the maximal trial duration was reached at least once in all conditions. We did not force a classification when the maximum trial duration was reached. Nevertheless, the classifier may still have been able to make a prediction, albeit at a lower confidence. To assess potential classifier behaviour under these circumstances, we performed a post-hoc analysis on the timed-out trials using the classifier state saved during the experiment. To stay consistent with the original approach, classification was based on the maximum window length extracted from the end of each timed-out trial.

For c-MVEP, $7.9$\,\% of all trials across participants reached the time limit. If classifications had been forced, an average accuracy of $87.0\,\%$ would have been reached. For c-VEP, only a single trial reached the maximum duration and was incorrectly classified in the post-hoc analysis, leaving the overall mean decoding accuracy unchanged at $97.8\,\%$.
For SSMVEP, $17.5\,\%$ of trials reached the maximum duration. Forced classification of these trials would have resulted in a mean accuracy of $69.9\,\%$ could have been reached. Lastly, for SSVEP, $2.2\,\%$ of trials timed-out, and forcing a classification would have resulted in a mean accuracy of $94.7\,\%$.

\begin{table}[h]
    \centering
    \caption{\textbf{Average Performance and SD.}  Mean and standard deviation values for all metrics (Accuracy, Selection Time, ITR) for each condition (c-MVEP, c-VEP, SSMVEP, SSVEP).}
    \label{tab:performance_table}
    \begin{tabular}{lccc}
        \hline\\[-1.5ex]
        \textbf{Condition} & \textbf{Accuracy} & \textbf{Selection Time} & \textbf{ITR} \\
        \hline\\[-1.5ex]        
        c-MVEP & $85.67 \pm 14.41$ & $2.61 \pm 1.09$ & $20.11 \pm 10.71$ \\
        c-VEP & $97.81 \pm 3.65$ & $1.15 \pm 0.37$ & $35.56 \pm 5.93$ \\
        SSMVEP & $64.91 \pm 23.21$ & $4.18 \pm 1.62$ & $10.13 \pm 10.53$ \\
        SSVEP & $93.42 \pm 9.18$  & $1.94 \pm 1.20$ & $28.12 \pm 9.78$ \\
        \hline
    \end{tabular}
\end{table}

\subsubsection{Subjective Rating}
\begin{figure}[h]
\centering
\includegraphics[width=\textwidth]{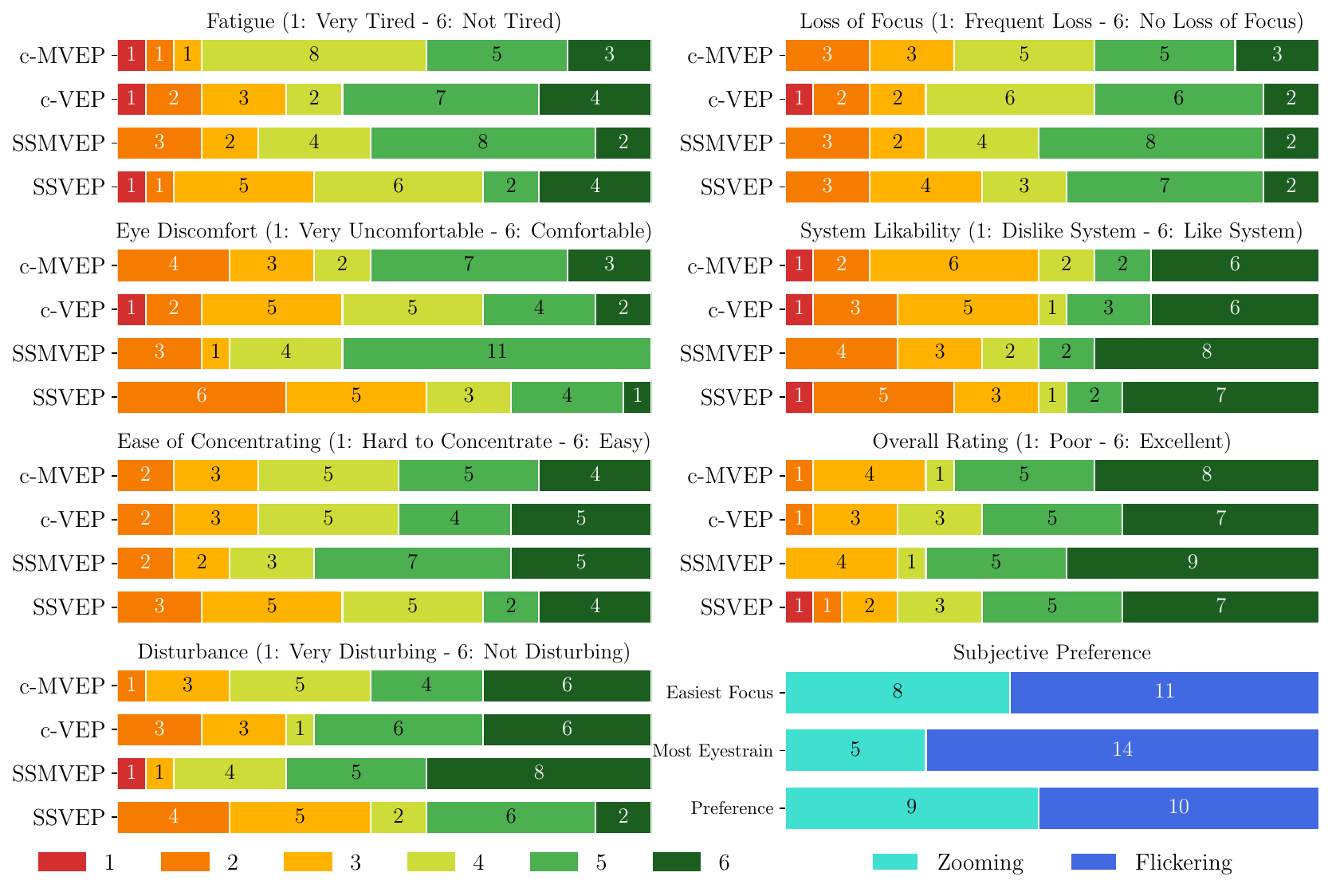}
\caption{\textbf{Results Online Questionnaire.} The answers to the questionnaire from the online experiment. Questions 1 - 7 were answered on a 6-point Likert scale. Here, 1 indicates a more negative rating, and 6 a more positive rating. The bottom-right bars depict the answers to the binary forced-choice preference questions (zooming or flickering). The numbers in the bars indicate the number of participants giving that specific score.}\label{fig:online_questionnaire}
\end{figure}

In the online experiment, the participants completed a questionnaire after each training session, prior to starting the testing session of a condition, to avoid the performance of the BCI influencing the answers. It consisted of seven questions assessing the participant's perceived fatigue level, visual discomfort of the stimulus, ease of concentration, disturbance, perceived loss of focus, system likability, and an overall rating. After completing all experimental conditions, participants were asked to compare the flickering and zooming stimulation types by indicating which was easiest to maintain focus on, which caused the most eye strain, and their final overall preference. The answers are shown in \autoref{fig:online_questionnaire}, with the average and standard deviation values in \autoref{tab:questionnaireOnline}. To maintain consistency across all subjective measures, the `Fatigue' and `Loss of Focus' items have been reverse-scored, such that a higher value indicates more favorable outcomes, aligning it with the other questions.

The non-parametric Friedman test was used to test for a significant difference.
No significant difference was found for the reported fatigue, discomfort rating, ease of concentration, loss of focus, system likability, and overall preference (all $p \geq .073$). 
A significant difference was found for the disturbance ($\chi^2 = 8.475$, $p=.037$); however, the post-hoc Wilcoxon signed-rank test with Bonferroni correction ($\alpha_{adj} = .0083$) revealed no significant differences between specific pairs of conditions (all $p \geq .020$).

Responses to the final questionnaire were analyzed separately, as these questions did not correspond to the individual conditions. These were binary yes/no questions and forced-choice preference questions (zooming or flickering). They were analyzed using two-sided binomial tests against the chance level ($p = .5$). Statistical significance was at $\alpha = 0.05$.
A large majority of the participants (17/19) reported perceiving differences in the flickering stimuli (c-VEP and SSVEP), which was significantly above chance level ($p < .001$). In contrast, although most participants (14/19) reported perceiving differences in the zooming stimuli (c-MVEP and SSMVEP), but this did not reach statistical significance ($p = .064$).
Lastly, no significant differences were found in the forced-choice preference questions regarding which was easiest to focus on (8 zooming, 11 flickering; $p = .648$), which caused the most eye strain (5 zooming, 14 flickering; $p = .064$), and for overall preference (9 zooming, 10 flickering; $p > .999$).

\begin{table}[]
\caption{\textbf{Questionnaire Results Online} Average (SD) results of the online questionnaire. Higher scores indicate more favourable ratings.}\label{tab:questionnaireOnline}
\setlength{\tabcolsep}{3pt}
\begin{tabular}{lccccccc} \hline\\[-1.5ex]
                & \textbf{Fatigue} & \textbf{Discomfort} & \textbf{Concentrating} & \textbf{Disturbance} & \textbf{Focus}  & \textbf{Likability} & \textbf{Rating} \\ \hline\\[-1.5ex] 
\textbf{c-MVEP} & $4.26$ $(1.25)$  & $4.11$ $(1.41)$  & $4.32$ $(1.26)$        & $4.58$ $(1.23)$      & $4.11$ $(1.29)$ & $4.05$ $(1.61)$     & $4.79$ $(1.32)$ \\
\textbf{c-VEP}  & $4.26$ $(1.48)$  & $3.79$ $(1.32)$  & $4.37$ $(1.31)$        & $4.47$ $(1.46)$      & $4.05$ $(1.32)$ & $4.05$ $(1.67)$     & $4.74$ $(1.25)$ \\
\textbf{SSMVEP} & $4.21$ $(1.24)$  & $4.21$ $(1.10)$  & $4.58$ $(1.27)$        & $4.89$ $(1.29)$      & $4.21$ $(1.24)$ & $4.37$ $(1.63)$     & $5.00$ $(1.17)$ \\
\textbf{SSVEP}  & $4.00$ $(1.38)$  & $3.42$ $(1.27)$  & $3.95$ $(1.36)$        & $3.84$ $(1.35)$      & $4.05$ $(1.28)$ & $4.00$ $(1.81)$     & $4.63$ $(1.46)$
\end{tabular}
\end{table}

\section{Discussion}
This study proposed a novel code-modulated motion VEP paradigm and systematically compared it against each of its directly related protocols, including c-VEP, SSMVEP, and SSVEP. 
We proposed the c-MVEP paradigm as a moving variant of the well-performing and robust c-VEP stimulation protocol, with the aim of providing a more comfortable yet equally reliable alternative. Specifically, our goal was to achieve the improved visual comfort associated with motion-stimulation while maintaining the robust performance advantages that the code-modulation holds over steady-state paradigms.
The study consisted of both an offline and online experiment to investigate the response characteristics of c-MVEP and to study the feasibility of the c-MVEP for BCI applications and the online BCI performance across all four paradigms. 

Overall, our results showed that the c-MVEP stimulation was able to evoke a code-modulated, broad-band response, similar to that observed in c-VEP.
We found a significant advantage of using c-MVEP over SSMVEP, in terms of accuracy, selection times, and ITR. However, both c-VEP and SSVEP outperformed c-MVEP and SSMVEP. From both subjective questionnaires, we did not find any statistical preference or comfort advantage for the motion paradigms. When flickering stimulation is not possible and motion stimulation is the preferred stimulation, the c-MVEP is a viable alternative, outperforming the current motion-based stimulation offered by SSMVEP.

\subsection{Experiment 1: Offline Response Analysis}
The offline experiment was used to descriptively investigate the response characteristics of c-MVEP compared to its three related paradigms, using an individual stimulus on the screen.
For c-MVEP and c-VEP, the grand-averaged evoked waveform (\autoref{fig:cvep_characteristics}) is non-trivial to interpret due to the core principles of the code-modulated paradigm, however, we may see that the motion caused different responses than the c-VEP flickering, although certain patterns might be recognized in both. 

In the frequency spectra, we observed that the c-MVEP evoked a broad-band response, similar to c-VEP, although it appeared more focused in the lower part of the frequency domain. In line with \citet{bin2009}, the responses evoked by c-MVEP and c-VEP contain multiple prominent frequencies, spread throughout the spectrum, confirming that both evoked the expected broad-band response. Striking in the c-MVEP plot, however, was that there appeared to be more variability between subjects than could be observed in c-VEP. 

The visibly lower amplitudes of the evoked waveform in c-MVEP and the overall larger inter-subject variability for both the time and frequency domain could be explained by the fact that some participants noted that it was hard to concentrate on the zooming stimuli, which may have influenced the response.

For the spatial distribution, we observed that c-MVEP appeared to be more spread around and away from the Oz electrode than c-VEP, which was visibly more focused around Oz. This was in line with what was observed in the SSMVEP domain, where SSVEP appeared mainly focused in the occipital region, and SSMVEP visibly spread out more towards the middle temporal vision region~\citep{han2018a}. 
Nonetheless, our analysis was limited to the distribution of the $20$\,Hz power, equal to the presentation rate of the m-sequence. Alternatively, the c-MVEP motion and c-VEP flicker responses could be characterized using all frequency components associated with the various event durations within the m-sequence.

The evoked waveform from SSMVEP and SSVEP (\autoref{fig:ssvep_characteristics}) was characterized by sinusoidal response, which was expected and similar to SSMVEP literature~\citep{han2018}. For SSMVEP, it appeared that the waveform had higher variability in subjects and did not synchronize at the stimulation frequency as SSVEP appeared to do. Both SSMVEP and SSVEP showed strong SNRs for the fundamental frequency and the harmonic frequencies. The harmonic responses were stronger for SSVEP than for SSMVEP, which is not in line with~\citet{stawicki2020}, who observed high harmonic responses for SSMVEP, nor in line with~\citet{han2018}, who reported only less obvious spectrum peaks for the second harmonic. 
Similar to what we observed for c-MVEP and c-VEP, here SSMVEP showed activity more spread away from Oz than SSVEP, where the activity was focused at Oz, in line with \citet{han2018a}.

From the SNR plots, it may be observed that adding motion to the SSVEP paradigm seems to result in a larger decrease in SNR than when adding motion to the c-VEP paradigm. 

For a practical BCI design, it should be taken into account that a zooming motion could result in decreased SNR, and more spread around Oz. Additionally, for c-MVEP, although a broad-band response have been observed, for decoding approaches that are based on specific events, such as the falling or rising edge of the m-sequence, it should be taken into account that the evoked waveform responses differed from those of c-VEP. It remains unclear whether the responses to these specific events are informative enough for reconvolution-based approaches~\citep{thielen2015, thielen2021}.

Since the offline study used one target, the observed response characteristics may not generalize to a multi-target BCI, in which multiple other nearby fields are presented simultaneously. The presence of these fields may influence the neural response to the target.

\subsection{Experiment 2: BCI Performance}
The online experiment investigated the feasibility of c-MVEP for online BCI applications and compared the protocol to its related stimulation paradigms. 
Overall, both flickering paradigms outperformed the motion-based paradigms, for accuracy, selection time, and ITR. Specifically, c-VEP was the best-performing protocol and reached significantly higher accuracies than c-MVEP and SSMVEP. c-VEP furthermore had the shortest selection times, significantly shorter than c-MVEP, SSMVEP, and SSVEP. Finally, c-VEP reached a significantly higher ITR than c-MVEP, SSMVEP, and SSVEP. 

Between c-MVEP and SSMVEP, it was found that c-MVEP reached significantly higher accuracies (c-MVEP: $85.67\%$, SSMVEP: $64.91\,\%$). It has to be noted here that the SSMVEP performance observed in the current study did not correspond to performances reported in~\citet{chai2019, chai2020, volosyak2020}, where accuracies of $89.7\,\%$, $94.9\,\%$, and $91.1\,\%$, respectively. The aforementioned studies used a CCA classification approach using sine and cosine templates, whereas the current study opted for CCA with user-specific templates. In addition, SSMVEP frequencies used in the current study were chosen to be close to the c-MVEP movement, and were also in line with previous studies~\citep{chai2019, chai2020, gao2019, gao2019a}, which again reported higher accuracies than observed in the current study. However, it is uncertain whether this led to the observed difference in performance or if other factors, such as stimulus size or zooming amplitude, contributed to this difference.

For selection times, a similar finding was observed, where c-MVEP ($2.61\pm1.09$\,s) was significantly faster than SSMVEP ($4.18\pm1.62$\,s). 
Similarly, c-MVEP reached significantly higher ITR than SSMVEP. 
When comparing this to c-VEP and SSVEP, a similar pattern was observed, where c-VEP was significantly faster in terms of selection time and reached significantly higher ITR. Only for accuracy, there was no significant difference.

Lastly, in the plots in \autoref{fig:performances_online_acc_selec}, there appeared to be more inter-subject variability in subjects for SSMVEP than c-MVEP. Similarly, SSVEP showed more inter-subject variability than c-VEP.
This would indicate that the code-modulated paradigm more consistently reached higher accuracies, faster selection times, and ITR than the frequency-based paradigm. Additionally, this might indicate that the frequency-based protocols rely more on optimization for individuals than the code-modulated protocols.

When investigating the effect of adding motion to c-VEP and SSVEP paradigms, it was observed that for the decoding accuracy, the reduction in accuracy and increase in selection times when going from SSVEP to SSMVEP was significantly larger than when going from c-VEP to c-MVEP. No such difference was observed for ITR. 
For both paradigms, it could be observed that, when adding motion, there is more subject variability.
Nonetheless, the code-modulated paradigm seemed more robust against the addition of motion into the paradigm than the frequency paradigm.

In line with our hypothesis, we found that the code-modulated paradigms performed significantly better in terms of selection time and ITR than the steady-state paradigms, in their respective protocol. For the motion-based paradigms, we additionally found that c-MVEP reached significantly higher decoding accuracies than SSMVEP, providing more reliable performances.

\subsection{Subjective Rating}
To assess user experience, subjective ratings were collected for all conditions in both the offline and online experiments.
For the offline experiment, participants answered the questionnaire following the first run of a given condition, whereas in the online experiment, the questions were presented after the training block but prior to the testing block, such that the BCI performance could not influence their responses. 
The questions were used to assess the level of comfort of the different stimulation conditions.
From both questionnaires, we found no preference for SSMVEP over SSVEP for comfort, which is not in line with results from subjective analysis in~\citet{chai2019}, who specifically showed a preference for the radial zooming stimulus over flickering. However, the results were in line with~\citet{volosyak2020}, where no statistical difference was found for comfort ratings between c-VEP, SSVEP, and SSMVEP.
Similarly, we found no preference for c-MVEP over the flickering c-VEP and SSVEP conditions. 

To assess subjective preference between the moving and flickering paradigms, participants completed a final questionnaire at the end of the online experiment. No significant differences were found between zooming and flickering for ease of focusing, causing most eye-strain or overall preference.
Contrary to our hypothesis that the motion-based stimulation would be more comfortable than the flickering stimulation, this study showed no clear benefit on subjective comfort for any of the conditions.

This was striking, since in the study we used a presentation rate of $20$\,Hz for c-MVEP and c-VEP, instead of the commonly used $60$\,Hz, and it was shown that c-VEP flickering at higher presentation rates was perceived as less fatiguing~\citep{martinez-cagigal2021}. Since the current study used a lower presentation rate, it would be expected that the motion stimulation was preferred over this flickering. 

The pre-experiment questionnaire of the online experiment asked about BCI experience, level of tiredness, and hours slept the previous night. Since these may have an influence on the participants' attention and BCI performance, the Supplementary Material includes jittered scatter plots fitted with a linear regression model, to visualize the relationship between the predictors and the c-MVEP average decoding performance. These exploratory analyses did not reveal any effect. We want to stress that a more in-depth analysis is needed to study the effect of fatigue on c-MVEP, especially since it was shown that SSMVEP decoding performance was less affected by fatigue than SSVEP~\citep{xie2016}. Future work should investigate if this also holds for c-MVEP and c-VEP.

\subsection{Limitations and Future Work}
The current findings show the feasibility of adding motion to the c-VEP protocol for BCIs. Nonetheless, further research is necessary to optimize the stimulation and decoding paradigm.

First, the current performances were found with a basic template-matching classifier, which was not optimized for any of the current protocols, but was still able to evoke high BCI performance. Future work should focus on improving the decoding by taking inspiration from decoding approaches of the c-VEP domain, such as the reconvolution-based CCA~\citep{thielen2021}, as well as the SSMVEP domain, such as task-related component analysis~\citep{nakanishi2017, stawicki2021}.
Especially when using the c-VEP optimized decoding approaches, we should investigate how to further adapt and optimize the analysis to suit the motion and the smoothed m-sequence. 

Second, it appears that c-MVEP would provide a better alternative for high-contrast flickering than SSMVEP in terms of decoding performance, especially for those who prefer the motion over the flickering stimuli. 
However, the amplitude of the zooming motion could be optimized. Specifically, here we opted for a $50$\,\% reduction in size, since too large zooming amplitudes could be perceived as uncomfortable and might cause a less smooth stimulation due to the limited number of frames available. On the other hand, a small zooming amplitude might be more comfortable, but might simultaneously not evoke a strong enough response. 
Additionally, the stimulus size for c-MVEP should be further investigated. For SSMVEP, \citet{chai2020} showed that the decoding performance of radial zooming with smaller stimuli was less affected by fatigue than the standard-sized stimulus. It remains unclear whether the visual comfort would also be improved with this change, or if this would hold for c-MVEP.
To sum up, the current implementation of motion could be improved by using a less invasive motion stimulation, for instance, by moving textures on the stimulus instead of the full stimulus itself. Such an improvement could provide comfort benefits of c-MVEP over flickering stimulation, despite the results found in the current sample.

Additionally, a lower zooming amplitude could alleviate the disadvantage of the currently used implementation. Specifically, the size of the stimulus matters for ERP response, creating a disadvantage for c-MVEP and SSMVEP compared to c-VEP and SSVEP, in the current study, since the latter two were always presented at $100$\,\% size. Additionally, when opting for a size change, we would advise having the center-point correspond to the size of the flickering stimuli, such that a $50$\,\% zoom corresponds to zooming between $75$\% and $125$\% to make the size component more fair.

A benefit of the motion stimuli is that, as shown by~\citet{gao2021a, gao2021b}, it is less affected by competing stimuli than the flickering stimuli. This could be a major advantage of c-MVEP as an alternative to c-VEP, compared to other alternatives proposed, such as the non-binary grey-flickering c-VEP~\citep{martinez-cagigal2023}.

The current work used a small offline sample size, due to the exploratory design of the study. Additionally, the results of this offline study were descriptive. A larger sample size should be considered for a fully quantitative offline EEG comparison between c-MVEP and related visual paradigms. 
In addition, this work used a four-target BCI application with horizontally neighboring targets, whereas many practical BCI applications aim for a higher target count with denser target layouts. Future work should investigate whether the current findings, both in terms of decoding performance and subjective ratings, will generalize to layouts with a larger number of targets.

During the online experiment, the gaze-shift interval was extended for three participants who explicitly reported that the gaze-shift period was too short. Since the gaze-shift duration is factored into the ITR, this extension affected their individual ITRs and likely contributed to the observed inter-subject variability. 
More broadly, however, this need for individual adaptation highlights a crucial point, where practical BCI applications require more individualized approaches. This applies not only to timing but also to the choice of stimulation protocol. Specifically, a customized BCI could use c-MVEP stimulation for users preferring the zooming stimulation and achieving high accuracies, while reserving traditional c-VEP for those who favor flickering stimuli. For practical BCI applications, we argue that the field should prioritize user-centered optimization, customizing the system parameters to individual needs and preferences, as was previously discussed for c-VEP~\citep{thielen2025} and SSVEP~\citep{siribunyaphat2025}.

Once the stimulation movement is improved and the decoding is more optimized for c-MVEP stimulation, the newly proposed c-MVEP may provide a good and reliable alternative to c-VEP, SSMVEP, and SSVEP.

\section{Conclusion}
This study presented, for the first time, the newly proposed code-modulated motion visual evoked potential (c-MVEP). With the new visual stimulus protocol, we attempted to combine the high performance of c-VEP with the improved comfort of SSMVEP. 
We performed a systematic comparison between c-MVEP, c-VEP, SSMVEP, and SSVEP, and found that c-MVEP outperformed SSMVEP in terms of accuracy, selection time, and ITR. However, the flickering stimulation outperformed the motion-based stimulation. 
There was no clear user preference for any of the conditions, that is, all conditions reached a similar comfort level.
In summary, we proposed a novel visual stimulus protocol, which could be tailored to individuals and provide a competitive alternative when others do not work or feel less comfortable for specific individuals. 
This study has taken the first step towards a flicker-free c-VEP BCI. By showing that c-MVEP can achieve reliable performance with comfort levels comparable to standard c-VEP and SSMVEP, we highlight its potential for creating effective, alternative BCI applications for users who prefer to avoid flickering stimuli.

\section*{Conflict of Interest Statement}
The authors declare that the research was conducted in the absence of any commercial or financial relationships that could be construed as a potential conflict of interest.

\section*{Author Contributions}
HS: Conceptualization, Methodology, Data Curation, Investigation, Formal Analysis, Visualization, Validation, Software, Writing - Original Draft, Writing - Review \& Editing; RH: Writing - Review \& Editing; JT: Conceptualization, Methodology, Writing - Review \& Editing; IV: Writing - Review \& Editing, Resources, Funding Acquisition, Supervision.

\section*{Funding}
This work was supported by the European Union's research and innovation programme under the Marie Sk\l{}odowska-Curie grant agreement No 101118964.

\section*{Acknowledgments}
The authors thank the financial support provided by ``The Friends of the University Rhine-Waal - Campus Cleve'' association. We also want to extend our appreciation to the participants who took part in this study.

\section*{Supplemental Data}

\section*{Data Availability Statement}
The datasets analyzed in this study are not currently available for public access. All EEG recordings were anonymized; however, our existing ethical approval does not cover public data sharing.

\section*{Ethics statement}
The studies involving humans were approved by the Ethics Committee of the Medical Faculty of the University of Duisburg-Essen, reference 24-11957-BO. All study procedures complied with applicable local legislation and institutional guidelines.

\bibliography{biblio}

\end{document}